\newif\ifonecol
\ifonecol

    \documentclass[12pt,draftclsnofoot,onecolumn]{IEEEtran}
\else
    \documentclass[journal,comsoc]{IEEEtran}
\fi

\usepackage{amsmath}
\usepackage[cmintegrals]{newtxmath}
\usepackage{bm}
\usepackage{algorithmic}
\usepackage{array}
\usepackage{algorithm}
\usepackage{cite}
\usepackage{multirow}
\usepackage{enumerate}
\usepackage{epsfig}
\usepackage[tight]{subfigure}
\usepackage{graphics}
\usepackage{color}

\newtheorem{thm}{Theorem}
\newtheorem{lem}{Lemma}
\newtheorem{cor}{Corollary}

\makeatletter
\newcommand{\vast}{\bBigg@{3.5}}
\newcommand{\Vast}{\bBigg@{7.5}}
\makeatother

\begin{document}
\ifonecol
    \title{\LARGE{Degrees of Freedom and Achievable Rate of Wide-Band Multi-cell Multiple Access Channels With No CSIT}}
\else
    \title{Degrees of Freedom and Achievable Rate of Wide-Band Multi-cell Multiple Access Channels With No CSIT}
\fi

\author{Yo-Seb Jeon, Namyoon Lee, and Ravi Tandon
\thanks{This paper was presented in part at the IEEE International Symposium on Information Theory (ISIT), June 2017 \cite{Jeon:17}.}
\thanks{Y.-S. Jeon and N. Lee are with the Department of Electrical Engineering, POSTECH, Pohang, Gyeongbuk, 37673 Korea (e-mail: yoseb.jeon@postech.ac.kr, nylee@postech.ac.kr).}
\thanks{R. Tandon is with the Department of Electrical and Computer Engineering, University of Arizona, Tucson, AZ 85721 USA (e-mail: tandonr@email.arizona.edu).}}
\vspace{-2mm}

\maketitle
\vspace{-12mm}

\begin{abstract}
This paper considers a $K$-cell multiple access channel with inter-symbol interference. The primary finding of this paper is that, without instantaneous channel state information at the transmitters (CSIT), the sum degrees-of-freedom (DoF) of the considered channel is $\frac{\beta -1}{\beta}K$ with $\beta \geq 2$ when the number of users per cell is sufficiently large, where $\beta$ is the ratio of the maximum channel-impulse-response (CIR) length of desired links to that of interfering links in each cell. Our finding implies that even without instantaneous CSIT, \textit{interference-free DoF per cell} is achievable as $\beta$ approaches infinity with a sufficiently large number of users per cell. This achievability is shown by a blind interference management method that exploits the relativity in delay spreads between desired and interfering links. In this method, all inter-cell-interference signals are aligned to the same direction by using a discrete-Fourier-transform-based precoding with cyclic prefix that only depends on the number of CIR taps. Using this method, we also characterize the achievable sum rate of the considered channel, in a closed-form expression.
% an illustrative example, we also show that when there exists a propagation delay between the desired and interfering links, the achievable sum-DoF is further improved by exploiting interference-free signals during the delay.
\end{abstract}


\section{Introduction}
 \IEEEPARstart{M}{ulti-cell} multiple access channel (MAC) with inter-symbol interference captures the communication scenario in which multiple uplink users per cell communicate with their associated base stations (BSs) by utilizing the same time and frequency resources across both the users and the BSs.
    The spectral efficiency of this channel is fundamentally limited by three different types of interference:
    \begin{itemize}
    	\item Inter-cell interference (ICI), which arises from simultaneous transmissions of users in neighboring cells;
    	\item Inter-user-interference (IUI), which is caused by simultaneous transmissions of multiple users in the same cell; and
    	\item Inter-symbol-interference (ISI), which occurs by the relativity between the transmit signal's bandwidth and the coherence bandwidth of a wireless channel.
    \end{itemize}

    Orthogonal frequency division multiple access (OFDMA) is the most well-known approach to mitigate both IUI and ISI in the multi-cell systems \cite{OFDM:66,Weinstein:71,Bingham:90}. The key idea of OFDMA is to decompose a wideband (frequency-selective) channel into multiple orthogonal narrowband (frequency-flat) subchannels, each with no ISI. By allocating non-overlapping sets of subchannels to the users in a cell, each user is able to send information data without suffering from ISI and IUI in the cell.  For instance, in a two-user MAC with ISI, which captures a single-cell uplink communication scenario, the capacity has been characterized by finding an optimal power allocation strategy across the subchannels \cite{Verdu:93,Wong:99,Rhee_Cioffi:00}.  These approaches, however, still suffer from ICI, which is a major hindrance towards increasing the spectral efficiency in multi-cell scenarios.

    Multi-cell cooperation has been considered as an effective solution to manage ICI problems for future cellular networks where BSs are densely deployed and small cells overlap heavily with macrocells \cite{Gesbert:10,Lee_Sayana:11,Clerckx_Kim:13}. The common idea of multi-cell cooperation is to form a BS cluster, which allows the information exchange among the BSs within the cluster, in order to jointly eliminate ICI. When multiple BSs in a cluster perfectly share the received uplink signals and channel state information (CSI) with each other via capacity-unlimited backhaul links, it is theoretically possible to eliminate ICI completely within the cluster. One problem with implementing multi-cell cooperation is that cooperation of  an entire network is not feasible considering the prohibited cost to establish high-capacity backhaul links.  As a practical solution, one could define multiple sets of BSs, multiple BS clusters, over an entire network, in which the BSs in a cluster are connected by high-capacity backhaul links. In this case, users (or BSs) outside the cluster are sources of interference, and this poses a fundamental limit to multi-cell cooperation even with capacity-unlimited backhaul links per cluster.  Another problem of multi-cell cooperation is that the amount of information that can be exchanged among BSs could be restricted due to capacity-limited backhual links. This possibly leads to the severe spectral efficiency loss that is caused by residual ICI.

    Among multi-cell cooperation strategies, coordinated beamforming provides a good tradeoff between the overheads for the information exchange and the gains on the spectral efficiency because this strategy only requires CSI exchange among the BSs in the same cluster \cite{Dahrouj_Yu:10,Lee_Heath:15}. Interference alignment (IA) is a representative coordinated beamforming method, which aligns ICI in a subspace so that the signal dimension occupied by interference is confined \cite{Cadambe_Jafar:08}. For example, in a single-input-single-output (SISO) interference channel,  IA has shown to be an optimal strategy in the sense of sum degrees-of-freedom (DoF) that characterizes the approximate sum-spectral efficiency in a high signal-to-noise-ratio (SNR) regime \cite{Cadambe_Jafar:08}.  The concept of IA has also been extended to multi-cell MACs (or interfering MACs) in single antenna settings \cite{Suh_Tse:08,Cadambe_Jafar:09} and multiple antenna settings \cite{Kim_Love:11,Lee_Shin:12,Yang_Paulraj:13}.  One remarkable result is that,  by an uplink IA method, the sum-DoF of $K$ is asymptotically achievable in the $K$-cell SISO MAC as the number of uplink users per cell approaches infinity \cite{Suh_Tse:08,Cadambe_Jafar:09,Chaaban:11}. The common requirement of prior works in \cite{Cadambe_Jafar:08,Suh_Tse:08,Cadambe_Jafar:09,Chaaban:11,Kim_Love:11,Lee_Shin:12,Yang_Paulraj:13} is global and perfect CSI at a transmitter (CSIT), which is a major obstacle in implementing these IA methods in practice.

	Recently, IA techniques using limited CSIT have extensively developed for various scenarios such as delayed CSIT \cite{Maddah-Ali:12}, mixed CSIT \cite{Gou_Jafar:12}, alternating CSIT \cite{Lee_Heath:12,Lee_Heath:14,Tandon:13}, one-bit CSIT \cite{Jafar_Index_coding}, and no CSIT \cite{Vaze:12,Vaze:12-2} (see the references therein \cite{Maddah-Ali:12,Gou_Jafar:12,Lee_Heath:12,Lee_Heath:14,Tandon:13,Jafar_Index_coding,Vaze:12,Vaze:12-2,Jafar:10}). Representatively, blind IA introduced by \cite{Wang_Jafar:11,Jafar:12,Jafar_Armada:15} has been considered as a practical IA technique when using limited CSIT. An attractive feature of blind IA is that it only requires to know autocorrelation functions of the channels in both time and frequency domains. This technique, however, heavily relies on the existence of the certain structure of channel coherence patterns, which may not be applicable for practical wireless environments in general.

    All the aforementioned multi-cell cooperation strategies have focused on the mitigation of ICI under the premise of perfect IUI and ISI cancellation by OFDMA. Recently, a blind interference management method has been proposed for the $K$-user SISO interference channel with ISI \cite{Lee:16}. The key idea of this method is to exploit the relativity of multi-path-channel lengths between desired and interfering links. This channel relativity allows the ICI alignment with discrete Fourier transform (DFT)-based precoding that needs no CSIT. One remarkable result in \cite{Lee:16} is that, without instantaneous CSIT, the sum-DoF of the $K$-user interference channel can be made to scale linearly with the number of communication pairs $K$, under some conditions on ISI channels.

\begin{figure}[t]
        \centering \vspace{-0.1cm}
        \epsfig{file=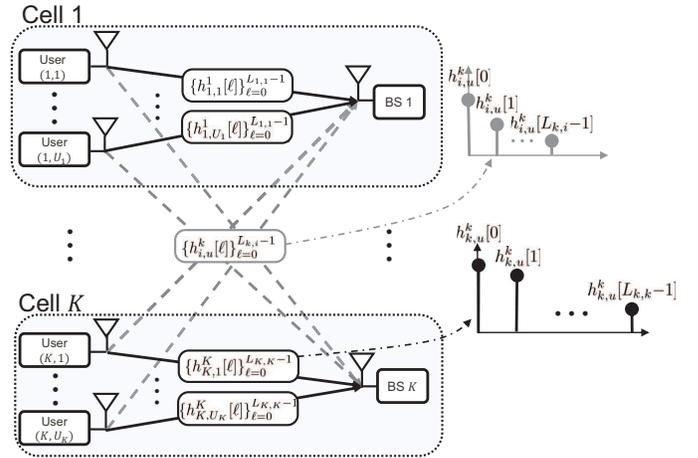, width=9cm}  \vspace{-0.1cm}
        \caption{An illustration of the system model for a $K$-cell MAC with ISI,  in which $U_k$ users are associated with the $k$-th BS.}
        \label{fig:Model}
    \end{figure}

    In this paper, we consider the $K$-cell SISO MAC with ISI, as illustrated in Fig.~\ref{fig:Model}. Continuing in the same spirit with \cite{Lee:16}, we attempt to characterize the sum-DoF of the multi-cell MAC with ISI in the absence of CSIT. Our major contribution is to demonstrate that, without instantaneous CSIT, the sum-DoF of the considered channel is
	     \begin{align}
	     	\left(1 - \frac{L_{\rm I}}{L_{\rm D}}\right)K,\nonumber
	     \end{align}
provided that the number of users per cell is larger than $L_{\rm D}-L_{\rm I}$ with $L_{\rm D} \geq 2L_{\rm I}$, where $L_{\rm D}$ and $L_{\rm I}$ are the maximum channel-impulse-response (CIR) lengths of desired and ICI links in each cell, respectively. Our result implies that \emph{interference-free DoF per cell} (i.e., sum-DoF of $K$) is achievable as $\frac{L_{{\rm D}}}{L_{\rm I}}$ approaches infinity with a sufficiently large number of users per cell. This result extends the sum-DoF result in \cite{Lee:16}, where the sum-DoF of $\frac{K}{2}$ is shown to be achievable without CSIT when each BS communicates with a single user. Therefore, our result also shows that communicating with multiple users in a cell provides a significant DoF gain compared to the single-user case, even in the absence of CSIT.
% This result extends the sum-DoF result in \cite{Lee:16}, where the sum-DoF of $\frac{K}{2}$ is shown to be achievable without CSIT under specific conditions on ISI channels when only a single user exists per cell. Therefore, we observe that communicating with multiple users in a cell provides a significant DoF gain when the number of users per cell is larger than the difference between the CIR taps of the desired and the interfering links.

    To demonstrate our result, we modify the blind interference management method in \cite{Lee:16}. The underlying idea of this method is to exploit the relativity in delay spreads between desired and ICI links. Specifically, by adding the cyclic prefix at transmitters (users) with an appropriate length and by removing it at receivers (BS), we create non-circulant matrices for the desired link, while generating circulant channel matrices for the ICI links.  This relativity in the matrix structure allows us to align all the ICI signals to the same direction by using a DFT-based precoding even in the absence of instantaneous CSIT, whereas making the desired signals spread over the entire signal dimensions.  As a result, all ICI can be simply canceled by using linear receive beamforming that does not depend on channel realizations.  After the ICI cancellation, each BS reliably decodes data symbols sent from the associated users by eliminating the remaining IUI and ISI perfectly, based on local CSI at a receiver (CSIR).

\vspace{3mm}
\subsubsection*{Notation}
Upper-case and lower-case boldface letters denote matrices and column vectors, respectively.
$\mathbb{E}[\cdot]$ is the statistical expectation, $\text{Pr}(\cdot)$ is the probability,
    $(\cdot)^\top$ is the transpose,
    $(\cdot)^H$ is the conjugate transpose,
    $\lceil \cdot \rceil$ is the ceiling function,
    $\lfloor \cdot \rfloor$ is the floor function,
    and $(x)^+ = \max\{x, 0\}$.
$\mathbf{I}_{m}$ is an $m\times m$ identity matrix,
    $\mathbf{1}_{m\times n}$ is an $m\times n$ all-one matrix,
    and $\mathbf{0}_{m\times n}$ is an $m$ by $n$ all-zero matrix.
$|\cdot|$ has three different meanings:
    $|a|$ denotes the absolute value of a scalar $a$;
    $|\mathcal{A}|$ denotes the cardinality of a set $\mathcal{A}$;
    and $|{\bf A}|$ denotes the determinant of a matrix ${\bf A}$,
%$\mathcal{CN}({\bf \mu},{\bf R})$ represents the distributions of a circularly-symmetric complex Gaussian random vector, which have the mean vector ${\bf \mu}$ and covariance matrix ${\bf R}$.

%%%%%%%%%%%%%%%%%% Fig. 1
%\begin{figure}[t]
%    \centering \vspace{-0.1cm}
%    \epsfig{file=Figures/Fig1_Model.eps, width=8.7cm}  \vspace{-0.1cm}
%    \caption{Illustration of a $3$-cell SISO MAC with ISI when two users are associated with a single BS in each cell.}
%    \label{fig:Model}
%\end{figure}

\section{System Model}
We consider a $K$-cell SISO MAC with ISI, where $U_k$ uplink users attempt to access a BS in cell $k$ for $k\in \mathcal{K}\triangleq \{1,2,\ldots,K\}$, by using the common time-frequency resources. We denote $\mathcal{U}_k$ as the index set of users associated with the $k$-th BS. Let $(k,u)$ be the user index denoting the $u$-th user in cell $k$. We assume that all users and BSs are equipped with a single antenna.  The CIR between a user (a transmitter) and a BS (a receiver) is represented by a finite number of channel taps.  We denote the CIR between user $(i,u)$ and the $k$-th BS by $\{h_{i,u}^k[\ell]\}_{\ell=0}^{L_{k,i}-1}$, where $h_{i,u}^k[\ell]$ is the $\ell$-th tap of the CIR, and $L_{k,i}$ is the number of the CIR taps. This length is typically defined as $L_{k,i} \triangleq \left\lceil T^{{\rm D},k}_{i,u} W_{\rm BW} \right\rceil$,
    where $W_{\rm BW}$ is the transmission bandwidth of the system, and $T^{{\rm D},k}_{i,u}$ is the delay spread of the wireless channel from user $(i,u)$ to the $k$-th BS.

We assume a block-fading channel model in which CIR taps are time-invariant during each block transmission. We also assume that each CIR tap, $h_{i,u}^k[\ell]$, is independently drawn from a continuous distribution for all $\ell \in \{1,\ldots,L_{k,i}\}$, $u \in \mathcal{U}_i$, and $i,k \in \mathcal{K}$. For example, in a rich-scattering propagation environment, CIR taps can be modeled as circularly symmetric complex Gaussian random variables.

Let $x_{k,u}[n]$ be the transmitted signal of user $(k,u)$ at time slot $n$ with the power constraint of $\mathbb{E}[|x_{k,u}[n]|^2]=P$.
Then the received signal of the $k$-th BS at time slot $n$ is
\begin{align}\label{eq:received}
    y_k[n]= \sum_{i=1}^K\sum_{u=1}^{U_i}\sum_{\ell=0}^{L_{k,i}-1}h_{i,u}^k[\ell]x_{i,u}[n-\ell]+ z_k[n],
\end{align}
where $z_k[n]$ is noise at the $k$-th BS in time slot $n$. We assume that $z_k[n]$ is independent and identically distributed (IID) circularly symmetric complex Gaussian random variable
    with zero mean and variance $\sigma^2$, i.e., $\mathcal{CN}(0,\sigma^2)$.

Throughout the paper, we assume no instantaneous CSIT, implying that all users (transmitters) do not have any knowledge of CIR taps $\{h_{i,u}^k[\ell]\}_{\ell=0}^{L_{k,i}-1}$ for $i\neq k$ and $i,k \in \mathcal{K}$. Furthermore, we assume that each BS is available to access knowledge of CSI between itself and the associated users in the cell, i.e., $\{h_{k,u}^k[\ell]\}_{\ell=0}^{L_{k,k}-1}$ for $k \in \mathcal{K}$. This is referred to as local CSIR. Note that local CSIR is necessary to perform coherent detection at the BSs.

%We provide some definitions for the ease of exposition.
%
%{\bf Definition~1 (User-index set):}
%Let $\mathcal{U}_k$ be the index set of users associated with the $k$-th BS. Furthermore, we define a subset $\mathcal{U}_{k,\ell} \subset \mathcal{U}_k$ as the collection of user index in the $k$th cell with the same CIR-length of $\ell$, i.e., $\mathcal{U}_{k,\ell} = \left\{u ~\Big| L_{k,u}^k = \ell \right\}$. Therefore, the user index set, $\mathcal{U}_k$, can be expressed as the union of the subsets, i.e., $\mathcal{U}_k=\bigcup_{\ell=1}^{L_{{\rm D},k}}\mathcal{U}_{k,\ell}$ where $L_{{\rm D},k} = \max_{u\in \mathcal{U}_k} \{L_{k,u}^k\}$.
%Note that by the definition, $\mathcal{U}_{k,1},\mathcal{U}_{k,2},\ldots,\mathcal{U}_{k,L_{{\rm D},k} }$ are mutually disjoint subsets of $\mathcal{U}_k$.

{\bf Definition (Sum degrees of freedom):}
User $(k,u)$ sends an independent message $m_{k,u}$ to the associated BS during $T$ time slots.
In this case, the rate of user $(k,u)$ is given by $R_{k,u}(P)=\frac{\log_2 |m_{k,u}|}{T}$.
The rate $\sum_{u=1}^{U_k}R_{k,u}(P)$ is \emph{achievable} if the $k$-th BS is able to decode the transmitted messages from the associating users
    with an arbitrarily-small error probability by choosing a sufficiently large $T$.
Then the \emph{sum-DoF}, which characterizes an approximate sum-spectral efficiency of the system at high SNR, is defined as
\begin{align}
    d_{\Sigma} & = \lim_{{P}\rightarrow \infty}\frac{\sum_{k=1}^{K} \sum_{u=1}^{U_k} R_{k,u}\left( P\right)}{\log\left({P}\right)}.
\end{align}

%%%%%%%%%%%%%%%%%% Fig. 2
\begin{figure*}[t]
    \centering \vspace{-0.1cm}
    \epsfig{file=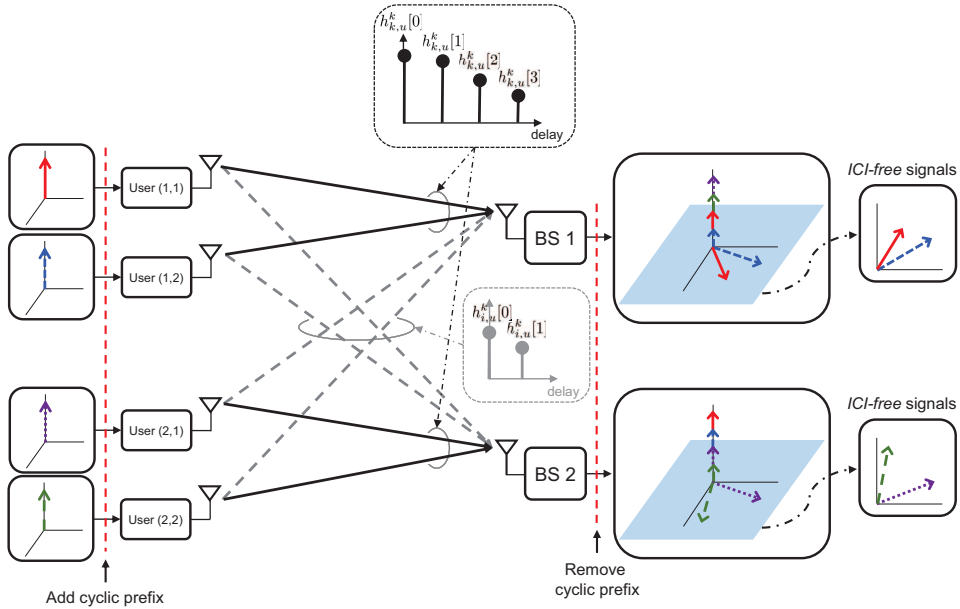, width=13cm}  \vspace{-0.1cm}
    \caption{An illustration of the conceptual process of the proposed interference management method for the two-cell MAC with ISI when two users exist per cell.}
    \label{fig:Concept}
\end{figure*}

\section{Blind Interference Management\\ using Channel Structural Relativity}
In this section, using a simple example scenario, we present the key concept of the proposed interference management method that exploits channel structural relativity. The generalization of this method will be presented in the sequel, to derive our main result.

\vspace{0.1cm}
{\bf Example 1 ($K$-cell MAC with Two Users):}
Consider a $K$-cell MAC with two users per cell. We assume a symmetric ISI case in which the channel-length of the desired links is four, i.e., $L_{k,k} =L_{\rm D}=4$ for $u\in \mathcal{U}_k$, while that of the ICI links is two, i.e., $L_{k,i} =L_{\rm I}=2$ for $u \in \mathcal{U}_i$, $i\neq k$, and $i, k\in\mathcal{K}\triangleq \{1,2,\ldots,K\}$. In this scenario, we show that it is possible to reliably decode total $2K$ data symbols with four time slots, i.e., $d_{\Sigma}=\frac{2K}{4}=\frac{K}{2}$, without CSIT.  The principal idea of the proposed interference management method is to exploit relativity in delay spreads between desired and ICI links to align all ICI signals to the same direction without instantaneous CSIT.

Let all users use a common precoding vector ${\bf f}_1=\frac{1}{\sqrt{3}}[1,1,1]^{\top}$ to send a data symbol. Then the precoded signal vector of user $(k,u)$, $\bar{\bf x}_{k,u} \in \mathbb{C}^3$, is
\begin{align}\label{eq:ex1:precoded}
    \bar{\bf x}_{k,u} = \big[\begin{array}{ccc}
            {x}_{k,u}[1],  & {x}_{k,u}[2], & {x}_{k,u}[3]
        \end{array}\big]^{\top} = {\bf f}_1 s_{k,u}.
\end{align}
By adding the cyclic prefix with the length of $L_{\rm I}-1 = 1$ to $\bar{\bf x}_{k,u}$ in \eqref{eq:ex1:precoded},
    the transmitted signal vector of user $(k,u)$ is
\begin{align}\label{eq:ex1:transmitted}
    {\bf x}_{k,u}
    &= \left[{\bf \bar x}_{k,u}^{\rm cp}, ~ {\bf \bar x}_{k,u} \right]^{\top} % \nonumber \\
    = \left[ {x}_{k,u}[3],  {x}_{k,u}[1] ,  {x}_{k,u}[2] ,  {x}_{k,u}[3] \right]^{\top}.
\end{align}
According to the above transmission strategy, every user uses four time slots to send one data symbol. The received signal vector of the $k$-th BS during four time slots are given by
\ifonecol

\vspace{-3mm}{\small{\begin{align}\label{eq:ex1:received}
    &\underbrace{\left[\begin{array}{c}
            {y}_{k}[1] \\ {y}_{k}[2] \\ {y}_{k}[3]  \\ {y}_{k}[4]  \\
        \end{array}\right]}_{{\bf y}_k}
    = \sum_{u=1}^2 \underbrace{\left[\begin{array}{cccc}
            {h}_{k,u}^k[0]  \!&\! 0              \!&\! 0              \!&\! 0 \\
            {h}_{k,u}^k[1]  \!&\! {h}_{k,u}^k[0] \!&\! 0              \!&\! 0 \\
            {h}_{k,u}^k[2]  \!&\! {h}_{k,u}^k[1] \!&\! {h}_{k,u}^k[0] \!&\! 0 \\
            {h}_{k,u}^k[3]  \!&\! {h}_{k,u}^k[2] \!&\! {h}_{k,u}^k[1] \!&\! {h}_{k,u}^k[0] \\
        \end{array}\right]}_{{\bf H}_{k,u}^k}  {\bf x}_{k,u}
     + \sum_{i\neq k}^K \sum_{u=1}^2 \underbrace{\left[
        \begin{array}{cccc}
            {h}_{i,u}^k[0]  \!&\! 0              \!&\! 0              \!&\! 0 \\
            {h}_{i,u}^k[1]  \!&\! {h}_{i,u}^k[0] \!&\! 0              \!&\! 0 \\
            0               \!&\! {h}_{i,u}^k[1] \!&\! {h}_{i,u}^k[0] \!&\! 0 \\
            0               \!&\! 0              \!&\! {h}_{i,u}^k[1] \!&\! {h}_{i,u}^k[0] \\
        \end{array}
        \right] }_{{\bf H}_{i,u}^k}
        {\bf x}_{i,u}
    + \underbrace{\left[\begin{array}{c}
            {z}_{k}[1] \\ {z}_{k}[2] \\ {z}_{k}[3]  \\ {z}_{k}[4]
        \end{array} \right]}_{{\bf z}_{k}}.
\end{align}}}\noindent
\else
\begin{align}\label{eq:ex1:received}
    &\underbrace{\left[\begin{array}{c}
            {y}_{k}[1] \\ {y}_{k}[2] \\ {y}_{k}[3]  \\ {y}_{k}[4]  \\
        \end{array}\right]}_{{\bf y}_k}
    = \sum_{u=1}^2 \underbrace{\left[\begin{array}{cccc}
            {h}_{k,u}^k[0]  \!&\! 0              \!&\! 0              \!&\! 0 \\
            {h}_{k,u}^k[1]  \!&\! {h}_{k,u}^k[0] \!&\! 0              \!&\! 0 \\
            {h}_{k,u}^k[2]  \!&\! {h}_{k,u}^k[1] \!&\! {h}_{k,u}^k[0] \!&\! 0 \\
            {h}_{k,u}^k[3]  \!&\! {h}_{k,u}^k[2] \!&\! {h}_{k,u}^k[1] \!&\! {h}_{k,u}^k[0] \\
        \end{array}\right]}_{{\bf H}_{k,u}^k}  {\bf x}_{k,u}
    \nonumber \\
    &\qquad + \sum_{i\neq k}^K \sum_{u=1}^2 \underbrace{\left[
        \begin{array}{cccc}
            {h}_{i,u}^k[0]  \!&\! 0              \!&\! 0              \!&\! 0 \\
            {h}_{i,u}^k[1]  \!&\! {h}_{i,u}^k[0] \!&\! 0              \!&\! 0 \\
            0               \!&\! {h}_{i,u}^k[1] \!&\! {h}_{i,u}^k[0] \!&\! 0 \\
            0               \!&\! 0              \!&\! {h}_{i,u}^k[1] \!&\! {h}_{i,u}^k[0] \\
        \end{array}
        \right] }_{{\bf H}_{i,u}^k}
        {\bf x}_{i,u}
    + \underbrace{\left[\begin{array}{c}
            {z}_{k}[1] \\ {z}_{k}[2] \\ {z}_{k}[3]  \\ {z}_{k}[4]
        \end{array} \right]}_{{\bf z}_{k}}.
\end{align}
\fi
%As mentioned previously, the key idea of interference-free OFDM is to 1) make all the channel matrices of interfering links ${\bf H}_{k,i}$ be circulant matrices, 2) while keeping desired channel matrix ${\bf H}_{k,k}$ not to hold the circulant matrix structure.
After removing the cyclic prefix from ${\bf y}_k$ in \eqref{eq:ex1:received}, the received signal vector of the $k$-th BS is obtained as
\ifonecol

\vspace{-3mm}{\small {\begin{align}\label{eq:ex1:cp_remove}
    &\underbrace{\left[\begin{array}{c}
            {y}_{k}[2] \\ {y}_{k}[3] \\ {y}_{k}[4] \\
        \end{array}\right]}_{{\bf \bar y}_k}
    = \sum_{u=1}^2 \underbrace{\left[%
        \begin{array}{ccc}
            {h}_{k,u}^k[0] \!&\! 0              \!&\! {h}_{k,u}^k[1] \\
            {h}_{k,u}^k[1] \!&\! {h}_{k,u}^k[0] \!&\! {h}_{k,u}^k[2]  \\
            {h}_{k,u}^k[2] \!&\! {h}_{k,u}^k[1] \!&\! {h}_{k,u}^k[0] {+} {h}_{k,u}^k[3] \\
        \end{array}
        \right]}_{{\bf \bar H}_{k,u}^k}  {\bf \bar x}_{k,u}
     + \sum_{i\neq k}^K \sum_{u=1}^2 \underbrace{\left[
        \begin{array}{ccc}
            {h}_{i,u}^k[0] \!&\! 0              \!&\! {h}_{i,u}^k[1] \\
            {h}_{i,u}^k[1] \!&\! {h}_{i,u}^k[0] \!&\! 0 \\
            0              \!&\! {h}_{i,u}^k[1] \!&\! {h}_{i,u}^k[0] \\
        \end{array}
        \right] }_{{\bf \bar H}_{i,u}^k}
        {\bf \bar x}_{i,u}
    + \underbrace{\left[\begin{array}{c}
            {z}_{k}[2] \\ {z}_{k}[3]  \\ {z}_{k}[4]
        \end{array} \right]}_{{\bf \bar z}_{k}}.
\end{align}}}\noindent
\else
\begin{align}\label{eq:ex1:cp_remove}
    &\underbrace{\left[\begin{array}{c}
            {y}_{k}[2] \\ {y}_{k}[3] \\ {y}_{k}[4] \\
        \end{array}\right]}_{{\bf \bar y}_k}
    = \sum_{u=1}^2 \underbrace{\left[%
        \begin{array}{ccc}
            {h}_{k,u}^k[0] \!&\! 0              \!&\! {h}_{k,u}^k[1] \\
            {h}_{k,u}^k[1] \!&\! {h}_{k,u}^k[0] \!&\! {h}_{k,u}^k[2]  \\
            {h}_{k,u}^k[2] \!&\! {h}_{k,u}^k[1] \!&\! {h}_{k,u}^k[0] {+} {h}_{k,u}^k[3] \\
        \end{array}
        \right]}_{{\bf \bar H}_{k,u}^k}  {\bf \bar x}_{k,u}
    \nonumber \\
    &\qquad + \sum_{i\neq k}^K \sum_{u=1}^2 \underbrace{\left[
        \begin{array}{ccc}
            {h}_{i,u}^k[0] \!&\! 0              \!&\! {h}_{i,u}^k[1] \\
            {h}_{i,u}^k[1] \!&\! {h}_{i,u}^k[0] \!&\! 0 \\
            0              \!&\! {h}_{i,u}^k[1] \!&\! {h}_{i,u}^k[0] \\
        \end{array}
        \right] }_{{\bf \bar H}_{i,u}^k}
        {\bf \bar x}_{i,u}
    + \underbrace{\left[\begin{array}{c}
            {z}_{k}[2] \\ {z}_{k}[3]  \\ {z}_{k}[4]
        \end{array} \right]}_{{\bf \bar z}_{k}}.
\end{align}
\fi
An important observation in \eqref{eq:ex1:cp_remove} is that all interference channel matrices ${\bf \bar H}_{i,u}^k$ for $i\neq k$ become circulant. Meanwhile, desired channel matrices ${\bf \bar H}_{k,u}^k$ can be expressed as a superposition of circulant and noncirculant matrices as follows:
\begin{align}\label{eq:ex1:cp_remove_cir}
    {\bf \bar H}_{k,u}^k = \underbrace{\left[\!\begin{array}{ccc}
            {h}_{k,u}^k[0] \!&\! {h}_{k,u}^k[2] \!&\! {h}_{k,u}^k[1] \\
            {h}_{k,u}^k[1] \!&\! {h}_{k,u}^k[0] \!&\! {h}_{k,u}^k[2] \\
            {h}_{k,u}^k[2] \!&\! {h}_{k,u}^k[1] \!&\! {h}_{k,u}^k[0] \\
        \end{array}
        \!\right]}_{{\bf \bar H}_{k,u}^{{\rm C}}}
    + \underbrace{\left[\!
        \begin{array}{ccc}
            0 \!&\! -{h}_{k,u}^k[2] \!\!&\!\! 0 \\
            0 \!&\! 0               \!\!&\!\! 0 \\
            0 \!&\! 0               \!\!&\!\! {h}_{k,u}^k[3] \\
        \end{array}
        \!\right]}_{{\bf \bar H}_{k,u}^{{\rm NC}}}.
\end{align}
Since a circulant matrix is diagonalized by DFT matrix, we can rewrite $ {\bf  \bar H}_{i,u}^k$ in \eqref{eq:ex1:cp_remove} and ${\bf \bar H}_{k,u}^{{\rm C}}$ in \eqref{eq:ex1:cp_remove_cir} as
\begin{align}
    {\bf  \bar H}_{i,u}^k
    =\left[\begin{array}{ccc}
        {\bf f}_1  & {\bf f}_2 & {\bf f}_3
    \end{array}\right]
    \left[\begin{array}{ccc}
        \lambda_{i,u,1}^k  \!\!&\!\!  0  \!\!&\!\!  0  \\
        0  \!\!&\!\!  \lambda_{i,u,2}^k  \!\!&\!\!  0  \\
        0  \!\!&\!\!  0  \!\!&\!\!  \lambda_{i,u,3}^k  \\
    \end{array}\right]
    \left[\begin{array}{ccc}
        {\bf f}_1 & {\bf f}_2 & {\bf f}_3
    \end{array}\right]^H,
\end{align}
and
\begin{align}
    {\bf  \bar H}_{k,u}^{\rm C}
    =\left[\begin{array}{ccc}
        {\bf f}_1  & {\bf f}_2 & {\bf f}_3
    \end{array}\right]
    \left[\begin{array}{ccc}
        \lambda_{k,u,1}^{\rm C}  \!\!&\!\!  0  \!\!&\!\!  0  \\
        0  \!\!&\!\!  \lambda_{i,u,2}^{\rm C}  \!\!&\!\!  0  \\
        0  \!\!&\!\!  0  \!\!&\!\!  \lambda_{i,u,3}^{\rm C}  \\
    \end{array}\right]
    \left[\begin{array}{ccc}
        {\bf f}_1 & {\bf f}_2 & {\bf f}_3
    \end{array}\right]^H,
\end{align}
where  ${\bf f}_n$ is the $n$-th column of 3-point IDFT matrix, and $\lambda_{i,u,n}^k$ and $\lambda_{k,u,1}^{\rm C}$ denote the $n$-th eigenvalues of ${\bf \bar H}_{i,u}^k$ and ${\bf \bar H}_{k,u}^{\rm C}$ associated with an eigenvector ${\bf f}_n$, respectively. As shown in \eqref{eq:ex1:precoded}, all users use the same precoding vector ${\bf f}_1$ when sending the data symbol, so the received signal vector ${\bf \bar y}_k$ is rewritten as
\begin{align}\label{eq:ex1:cp_remove2}
    &{\bf \bar y}_k
    = \sum_{u=1}^{2}{\bf \bar H}_{k,u}^k{\bf f}_1s_{k,u} + \sum_{i\neq k}\sum_{u=1}^{2}{\bf \bar H}_{i,u}^k{\bf f}_1s_{i,u}
        + {\bf \bar z}_{k} \nonumber \\
    &\!\!= \sum_{u=1}^{2}{\bf \bar H}_{k,u}^{\rm NC}{\bf f}_1s_{k,u}
        + \left\{\sum_{u=1}^{2}\lambda_{k,u,1}^{\rm C}s_{k,u} + \sum_{i\neq k}\sum_{u=1}^{2} \lambda_{i,u,1}^{k} s_{i,u} \right\}{\bf f}_1
        + {\bf \bar z}_{k}.
\end{align}
From \eqref{eq:ex1:cp_remove2}, one can easily see that all ICI signals from other-cell users are aligned in the same direction of ${\bf f}_1$. Therefore, it is possible to eliminate all the aligned ICI signals by multiplying an orthogonal projection matrix ${\bf W}= \left[{\bf f}_2, {\bf f}_3\right]^H$ to ${\bf \bar y}_k$. After the ICI cancellation, an effective received signal vector is obtained as
\ifonecol
\begin{align}\label{eq:ex1:effreceived}
    {\bf \tilde y}_k
    &= {\bf W} {\bf \bar y}_k
    = \sum_{u=1}^{2} {\bf W}{\bf \bar H}_{k,u}^{\rm NC}{\bf f}_1s_{k,u} + {\bf W}{\bf\bar z}_{k} = \underbrace{\left[\begin{array}{cc}
            {\bf f}_2^H{\bf \bar H}_{k,1}^{\rm NC}{\bf f}_1 & {\bf f}_2^H{\bf \bar H}_{k,2}^{\rm NC}{\bf f}_1 \\
            {\bf f}_3^H{\bf \bar H}_{k,1}^{\rm NC}{\bf f}_1 & {\bf f}_3^H{\bf \bar H}_{k,2}^{\rm NC}{\bf f}_1 \\
        \end{array}\right]}_{\tilde{\bf H}_k}
        \underbrace{\left[\begin{array}{c}
            s_{k,1} \\ s_{k,2}
        \end{array}\right]}_{{\bf s}_k}
        + \underbrace{\left[\begin{array}{c}
        {\bf f}_2^H{\bf\bar z}_{k} \\ {\bf f}_3^H{\bf\bar z}_{k}
    \end{array}\right]}_{\tilde{\bf  z}_k}.
\end{align}
\else
\begin{align}\label{eq:ex1:effreceived}
    {\bf \tilde y}_k
    &= {\bf W} {\bf \bar y}_k
    = \sum_{u=1}^{2} {\bf W}{\bf \bar H}_{k,u}^{\rm NC}{\bf f}_1s_{k,u} + {\bf W}{\bf\bar z}_{k} \nonumber \\
    &= \underbrace{\left[\begin{array}{cc}
            {\bf f}_2^H{\bf \bar H}_{k,1}^{\rm NC}{\bf f}_1 & {\bf f}_2^H{\bf \bar H}_{k,2}^{\rm NC}{\bf f}_1 \\
            {\bf f}_3^H{\bf \bar H}_{k,1}^{\rm NC}{\bf f}_1 & {\bf f}_3^H{\bf \bar H}_{k,2}^{\rm NC}{\bf f}_1 \\
        \end{array}\right]}_{\tilde{\bf H}_k}
        \underbrace{\left[\begin{array}{c}
            s_{k,1} \\ s_{k,2}
        \end{array}\right]}_{{\bf s}_k}
        + \underbrace{\left[\begin{array}{c}
        {\bf f}_2^H{\bf\bar z}_{k} \\ {\bf f}_3^H{\bf\bar z}_{k}
    \end{array}\right]}_{\tilde{\bf  z}_k}.
\end{align}
\fi
Note that this effective received signal only contains the desired signals from the users in the own cell. The effective channel matrix $\tilde{\bf H}_k$ in \eqref{eq:ex1:effreceived} is decomposed as
    \begin{align}
        \tilde{\bf H}_k = -\frac{1}{3} \left[\begin{array}{cc}
            1 ~& \frac{1}{2}-\frac{\sqrt{3}}{2}j \\
            1 ~& \frac{1}{2}+\frac{\sqrt{3}}{2}j \\
        \end{array}\right]
        \left[\begin{array}{cc}
            {h}_{k,1}^k[2] & {h}_{k,2}^k[2] \\
            {h}_{k,1}^k[3] & {h}_{k,2}^k[3] \\
        \end{array}\right].
    \end{align}
From the above decomposition, we can easily show that $\text{rank}(\tilde{\bf H}_k)=2$ with probability one as $h_{k,u}^k[\ell]$ is independently drawn from a continuous distribution. Consequently, the data symbols sent from the two users, $s_{k,1}$ and $s_{k,2}$, are decodable at the $k$-thy BS by applying maximum-likelihood detection (MLD) or zero-forcing detection (ZFD) methods. Since four time slots are used to deliver $2K$ independent data symbols, the sum-DoF is given by $d_{\Sigma}=\frac{K}{2}$.

{\bf Remark 1 (The design of precoding vectors and projection matrices):}
One of important observations is that the proposed interference management method does not require instantaneous CSIT in the design of precoding vectors. As shown in the example, by judiciously choosing the length of the cyclic prefix, all channel matrices of ICI links become circulant. Therefore, to align ICI signals, we simply select an eigenvector of the circulant matrix (i.e., a column vector of the IDFT matrix) as the common precoding vector ${\bf f}_1$ regardless of channel realizations. This allows us to cancel ICI signals with no CSIT: By constructing orthogonal projection matrix ${\bf W}$ that are solely determined by ${\bf f}_1$, all the ICI signals are perfectly removed.  After the ICI cancellation, each BS is able to decode data symbols sent from the associated users only using local CSIR.

%%%%%%%%%%%%%%%%%%%%%%%%%%%%%%%%%%%%
\section{Main Result}
In this section, by generalizing the concept of the blind interference management method introduced in Section III, we establish the following result.
\vspace{0.2cm}
\begin{thm} \label{Theorem1}
    Consider a $K$-cell MAC with ISI, each cell with $U_k$ uplink users.
    Let $L_{\rm I} = \max_k\max_{i\neq k}L_{k,i}$ and $L_{\rm D} = \max_k L_{k,k}$.
    The achievable sum-DoF of this channel without instantaneous CSIT is
    \begin{align}\label{eq:thm1:sumDoF}
        d^{\rm MAC}_{\Sigma}  = \max\left\{ \sum_{k=1}^K  \frac{ \min \left\{U_kM_k,(L_{k,k}-L_{\rm I})^+\right\} }
        { \max\left\{ L_{\rm D} - L_{\rm I} + \max_k M_k, L_{\rm I}\right\}+L_{\rm I}-1},
        1 \right\},
    \end{align}
    where $M_k = \max\left\{\left\lfloor\frac{L_{k,k}-L_{\rm I}}{U_k}\right\rfloor, 1\right\}$.
\end{thm}
\vspace{0.3cm}
\begin{IEEEproof}
In this proof, we only focus on the case that
    \begin{align}
         \sum_{k=1}^K \! \frac{ \min \left\{U_kM_k,(L_{k,k}-L_{\rm I})^+\right\} }
        { \max\left\{ L_{\rm D} - L_{\rm I} + \max_k M_k, L_{\rm I}\right\}+L_{\rm I}-1} > 1,
    \end{align}
because otherwise, the trivial sum-DoF of one is achievable by using time-division multiple access (TDMA) among the BSs with OFDMA in each cell.

%%%%%%%%%%%%%%%%%%%%%%%%%%%%%%%%%%%%%%%%%
\vspace{0.2cm}
\textbf{Block transmission strategy:}
The key idea of this proof is similar to \cite{Lee:16} using a block transmission method. We start by presenting a block transmission strategy that consists of $B$ subblock transmissions.

During the block transmission in cell $k$, only $U_k^\prime \leq U_k$ users are active and each active user sends $M_k$ data symbols for each subblock.
% Let $M_k$ be the number of transmitted data symbols per user and $U_k^\prime \leq U_k$ be the number of active users during a subblock transmission in cell $k$.
Specifically, we determine $M_k$ and $U_k^\prime$ as follows:
\begin{itemize}
  \item When $U_k \leq L_{k,k}-L_{\rm I}$, we choose $U_k^\prime = U_K$ and $M_k = \left\lfloor\frac{L_{k,k}-L_{\rm I}}{U_k}\right\rfloor$,
    which implies that all the users transmit the $M_k$ data symbols.
  \item When $0 < L_{k,k}-L_{\rm I} < U_k$, we choose $U_k^\prime = L_{k,k}-L_{\rm I}$ and $M_k = 1$, which implies that randomly selected $U_k^\prime $ users among $U_k$ users transmit a single data symbol.
  \item When $L_{k,k}-L_{\rm I} \leq 0$, we choose $U_k^\prime = 0$ and $M_k = 0$, which implies that none of the users transmit data symbols.
\end{itemize}
%Then the total number of data symbols transmitted during a subblock transmission in cell $k$ is obtained as $\bar{M}_k = U_k^\prime M_k$.
Each subblock transmission consists of ${\bar N}=N + L_{\rm I}-1$ time slots, where $N =\max\{L_{\rm D}-L_{\rm I}+M_{\rm D}, L_{\rm I}\}$ and $M_{\rm D} = \max_k \{M_k\}$.
After transmitting $B$ subblocks, we append $\max\{L_{\rm D},L_{\rm I}\}-1$ zeros at the end of the transmission block, to avoid inter-block interference between two subsequent block transmissions. Therefore, the total number of time slots needed for a single block transmission is $T=B{\bar N} + \max\{L_{\rm D},L_{\rm I}\}-1$.
%In the sequel, we start from demonstrating how each cell achieves DoF of $d_{k}=\frac{B(L_{{\rm D},k} - L_{\rm I})}{B(N +L_{\rm I}-1)+ L_{\rm D}-1}$
%    when the number of subblocks $B$ is fixed, for the given user set $\mathcal{U}_k$ that satisfies the condition \eqref{eq:thm1:cond}.
%We present a  transmission method by leveraging a cyclic prefix based OFDM technique,
% as it is a more systematic transmission method than the one explained in Section II.

During the subblock transmission, each user $(k,u)$ transmits $M_k$ data symbols using a DFT-based precoding with cyclic fix. Let ${\bf s}_{k,u}^b=[s_{k,u,1}^b,s_{k,u,2}^b,\cdots,s_{k,u,M_k}^b] \in \mathbb{C}^{M_k}$ be the data symbol vector of user $(k,u)$ transmitted during the $b$-th subblock transmission. User $(k,u)$ uses a precoding matrix ${\bf F}_{k} = [{\bf f}_1,{\bf f}_2,\ldots,{\bf f}_{M_k}]\in\mathbb{C}^{N \times M_k}$ to send the data symbol vector ${\bf s}_{k,u}^b$, so the precoded data symbol vector, namely ${\bf \bar x}^b_{k,u}\in \mathbb{C}^{N}$, is given by
\ifonecol
    \begin{align}\label{eq:precoded}
        {\bf \bar x}^b_{k,u} &= \Big[x_{k,u}[(b-1){\bar N}+1], \cdots, x_{k,u}[(b-1){\bar N}+N]\Big]^{\!\top} = {\bf F}_k {\bf s}_{k,u}^b = \sum_{m=1}^{M_k} {\bf f}_{m}s_{k,u,m}^b,
    \end{align}
\else
    \begin{align}\label{eq:precoded}
        {\bf \bar x}^b_{k,u} &= \Big[x_{k,u}[(b-1){\bar N}+1], \cdots, x_{k,u}[(b-1){\bar N}+N]\Big]^{\!\top} \nonumber\\
        &= {\bf F}_k {\bf s}_{k,u}^b = \sum_{m=1}^{M_k} {\bf f}_{m}s_{k,u,m}^b,
    \end{align}
\fi
for $u \in \mathcal{U}_k$, $k \in \mathcal{K}$, and $b\in\{1,2,\ldots,B\}$. We add a cyclic prefix with length $L_{{\rm I}}-1$ to ${\bf \bar x}_{k,u}^{b}$ and thus generate the transmitted signal vector:
\begin{align}\label{eq:transmitted}
    {\bf x}_{k,u}^b = \left[{\bf \bar x}_{k,u}^{b,{\rm cp}}, ~ {\bf \bar x}_{k,u}^{b} \right]^{\!\top}  \in \mathbb{C}^{\bar N},
\end{align}
where
\ifonecol
    \begin{align}
        {\bf \bar x}_{k,u}^{b,{\rm cp}} = \left[x_{k,u}[(b{-}1){\bar N}+N-L_{\rm I}+2], \cdots, x_{k,u}[(b{-}1){\bar N}+N]\right]^{\!\top}.
    \end{align}
\else

    \vspace{-3mm}{\small\begin{align}
        {\bf \bar x}_{k,u}^{b,{\rm cp}} = \left[x_{k,u}[(b{-}1){\bar N}+N-L_{\rm I}+2], \cdots, x_{k,u}[(b{-}1){\bar N}+N]\right]^{\!\top}.
    \end{align}}
\fi

From the above strategy, the signal vector of user $(k,u)$ transmitted during a single block transmission is
\begin{align}
    {\bf x}_{k,u} = \Big[\left({\bf  x}_{k,u}^{1}\right)^{\!\top}\!\!,\left({\bf  x}_{k,u}^{2}\right)^{\!\top}\!\!,\cdots,\left({\bf  x}_{k,u}^{B}\right)^{\!\top}\!\!,\!\!\!
        \underbrace{0,\ldots,0}_{\max\{L_{\rm D},L_{\rm I}\}-1}\!\!\!\!\! \Big]^{\!\top}.
\end{align}
% Therefore, the total number of time slots needed for a single block transmission is $T=B{\bar N} + \max\{L_{\rm D},L_{\rm I}\}-1$.

%%%%%%%%%%%%%%%%%%%%%%%%%%%%%%%%%%%%%%%%%
\vspace{0.2cm}
\textbf{Received signal representation:}
We specify received signals at the BS during each subblock transmission. From \eqref{eq:received}, the received signal of the $k$-th BS at time slot $n$ of the $b$-th subblock transmission is represented as
\ifonecol
\begin{align}\label{eq:received_b}
    y_{k}[(b-1){\bar N} + n]
        = &\sum_{u=1}^{U_k^\prime}\sum_{\ell=0}^{L_{k,k}-1} h_{k,u}^k[\ell]x_{k,u}[(b-1){\bar N} + n -\ell] \nonumber \\
    & + \sum_{i=1,i\neq k}^{K} \sum_{u=1}^{U_i^\prime} \sum_{\ell=0}^{L_{k,i}-1} h_{i,u}^k [\ell]x_{i,u}[(b-1){\bar N}+n-\ell] + z_k[(b-1){\bar N} + n],
\end{align}
\else
\begin{align}\label{eq:received_b}
    &y_{k}[(b-1){\bar N} + n]
        = \sum_{u=1}^{U_k^\prime}\sum_{\ell=0}^{L_{k,k}-1} h_{k,u}^k[\ell]x_{k,u}[(b-1){\bar N} + n -\ell] \nonumber \\
    &\qquad\qquad~~~ + \sum_{i=1,i\neq k}^{K} \sum_{u=1}^{U_i^\prime} \sum_{\ell=0}^{L_{k,i}-1} h_{i,u}^k [\ell]x_{i,u}[(b-1){\bar N}+n-\ell]  \nonumber \\
    &\qquad\qquad~~~ + z_k[(b-1){\bar N} + n],
\end{align}
\fi
for $n\in\{1,2,\ldots, {\bar N}\}$ and $b\in\{1,2,\ldots, B\}$.
%Let ${\bf \bar y}_k^b= \big[y_k[(b-1){\bar N}+L_{\rm I}],\cdots,y_k[(b-1){\bar N}+{\bar N}] \big]^{\!\top}\in\mathbb{C}^N$ be the received signal of the $k$-th BS during the $b$-th subblock transmission after discarding the cyclic prefix, respectively.
After discarding the cyclic prefix of length $L_{\rm I}-1$, the received signal vector of the $k$-th BS during the $b$-th subblock transmission, namely ${\bf \bar y}_k^b \in\mathbb{C}^N$, is
\ifonecol
    \begin{align}\label{eq:cp_remove}
        {\bf \bar y}_k^b
        &= \Big[y_k[(b-1){\bar N}+L_{\rm I}],\cdots,y_k[(b-1){\bar N}+{\bar N}] \Big]^{\!\top} \nonumber \\
        &= \sum_{u=1}^{U_k^\prime}{\bf \bar H}_{k,u}^{k}{\bf \bar x}^b_{k,u}
            + \sum_{i=1,i\neq k}^K\sum_{u=1}^{U_i^\prime} {\bf \bar H}_{i,u}^{k}{\bf \bar x}^b_{i,u} + {\bf \bar z}_k^b
            = \sum_{u=1}^{U_k^\prime}{\bf \bar H}_{k,u}^k {\bf F}_{k} {\bf s}_{k,u}^b
            + \sum_{i=1,i\neq k}^K\sum_{u=1}^{U_i^\prime} {\bf \bar H}_{i,u}^k {\bf F}_{k} {\bf s}_{i,u}^b + {\bf \bar z}_k^b,
	\end{align}
\else
    \begin{align}\label{eq:cp_remove}
        {\bf \bar y}_k^b
        &= \Big[y_k[(b-1){\bar N}+L_{\rm I}],\cdots,y_k[(b-1){\bar N}+{\bar N}] \Big]^{\!\top} \nonumber \\
        &= \sum_{u=1}^{U_k^\prime}{\bf \bar H}_{k,u}^{k}{\bf \bar x}^b_{k,u}
            + \sum_{i=1,i\neq k}^K\sum_{u=1}^{U_i^\prime} {\bf \bar H}_{i,u}^{k}{\bf \bar x}^b_{i,u} + {\bf \bar z}_k^b \nonumber \\
        &= \sum_{u=1}^{U_k^\prime}{\bf \bar H}_{k,u}^k {\bf F}_{k} {\bf s}_{k,u}^b
            + \sum_{i=1,i\neq k}^K\sum_{u=1}^{U_i^\prime} {\bf \bar H}_{i,u}^k {\bf F}_{k} {\bf s}_{i,u}^b + {\bf \bar z}_k^b,
	\end{align}
\fi	
where ${\bf \bar H}_{i,u}^{k} \in \mathbb{C}^{N\times N}$ is a matrix representation of the convolution involved with ${\bf \bar x}^b_{i,u}$, and ${\bf \bar z}_k^b= \big[z_k[(b{-}1){\bar N}+L_{\rm I}],\cdots,z_k[(b{-}1){\bar N}+{\bar N} \big]^{\!\top} \in\mathbb{C}^N$ is the noise vector received by ${\bf \bar y}_k^b$.

For the ease of exposition, we define ${\rm Circ}({\bf c})$ as an $n$ by $n$ circulant matrix when its first column is ${\bf c} \in \mathbb{C}^n$. Using this notation, the channel matrix of the  ICI link between user $(i,u)$ and the $k$-th BS for $i\neq k$ is represented as
    \begin{align}\label{eq:interferenceH}
        {\bf \bar H}_{i,u}^{k} = {\rm Circ}\Big( \Big[{h}_{i,u}^k[0],\cdots,{h}_{i,u}^k[L_{k,i}{-}1],\underbrace{0,\ldots,0}_{N-L_{k,i}} \Big] \Big).
    \end{align}
Whereas, the channel matrix of the desired link between user $(k,u)$ and the $k$-th BS is decomposed into two matrices:
    \begin{align}\label{eq:desiredH}
        {\bf \bar H}_{k,u}^{k} = {\bf \bar H}_{k,u}^{\rm C} + {\bf \bar H}_{k,u}^{\rm NC},
    \end{align}
    where
    \begin{align}\label{eq:circulantH}
        {\bf \bar H}_{k,u}^{\rm C} =
            {\rm Circ}\Big( \Big[{h}_{k,u}^k[0],\cdots,{h}_{k,u}^k[N_{k,u}^\prime{-}1],\underbrace{0,\ldots,0}_{(N-L_{k,k})^+} \Big] \Big),
    \end{align}
    with $N_{k}^\prime = \min\{L_{k,k},N\}$,
    and ${\bf \bar H}_{k,u}^{\rm NC}$ has the form of
    \begin{align}\label{eq:noncirculantH}
        {\bf \bar H}_{k,u}^{\rm NC} = \left[\begin{array}{cc}
            -{\bf \bar H}_{k,u}^{\rm upp}  &  {\bf 0}_{(N-L_{\rm I}+1) \times (L_{\rm I}-1)} \\
            {\bf 0}_{(L_{\rm I}-1) \times (N-L_{\rm I}+1)} &  {\bf \bar H}_{k,u}^{\rm low}  \\
        \end{array}\right].
    \end{align}
In \eqref{eq:noncirculantH}, ${\bf \bar H}_{k,u}^{\rm upp} \in \mathbb{C}^{(N-L_{\rm I}+1) \times (N-L_{\rm I}+1)}$ and ${\bf \bar H}_{k,u}^{\rm low}\in \mathbb{C}^{(L_{\rm I}-1) \times (L_{\rm I}-1)}$ are upper and lower toeplitz matrices defined in \eqref{eq:toepH} (see the top of the page).
\ifonecol
    \begin{figure*}
    {\scriptsize{\begin{align}\label{eq:toepH}
        {\bf \bar H}_{k,u}^{\rm upp} = \left[\begin{array}{cccccc}
        0      &~~\cdots~~  & 0      &\!\!\! h_{k,u}^k[N_{k}^\prime{-}1] &\!\! \cdots &\!\! h_{k,u}^k[L_{\rm I}] \\
        \vdots & 0          & ~      &\!\!\! 0                                &\!\! \ddots &\!\! \vdots \\
        \vdots & ~          & \ddots &\!\!\! ~                                &\!\! \ddots &\!\! h_{k,u}^k[N_{k}^\prime{-}1] \\
        \vdots & ~          & ~      &\!\!\! \ddots                           &\!\! ~      &\!\! 0 \\
        \vdots & ~          & ~      &\!\!\! ~                                &\!\! \ddots &\!\! \vdots \\
        0      & 0          & \cdots &\!\!\! \cdots                           &\!\! \cdots &\!\! 0 \\
        \end{array}\right]~~\text{and}~~
        {\bf \bar H}_{k,u}^{\rm low} = \left[\begin{array}{cccccc}
        h_{k,u}^k[N]                \!\!\!\!\!\!&\!\!\!\!\!\! 0             & \cdots   &\!\!\! \cdots                     \!\!\!& \cdots & 0 \\
        \vdots                      \!\!\!\!\!\!&\!\!\!\!\!\! h_{k,u}^k[N]  & 0        &\!\!\! ~                          \!\!\!& ~      & \vdots \\
        h_{k,u}^k[L_{k,k}{-}1]  \!\!\!\!\!\!&\!\!\!\!\!\! ~             & \ddots   &\!\!\! \ddots                     \!\!\!& ~      & \vdots \\
        0                           \!\!\!\!\!\!&\!\!\!\!\!\! \ddots        & ~        &\!\!\! \ddots                     \!\!\!& \ddots & \vdots \\
        \vdots                      \!\!\!\!\!\!&\!\!\!\!\!\! \ddots        & \ddots   &\!\!\! ~                          \!\!\!& \ddots & 0 \\
        0                           \!\!\!\!\!\!&\!\!\!\!\!\! \cdots        & 0        &\!\!\! h_{k,u}^k[L_{k,k}{-}1] \!\!\!& \cdots & h_{k,u}^k[N] \\
        \end{array}\right].
    \end{align}}}
    \hrulefill
    \end{figure*}
\else
    \begin{figure*}
    \begin{align}\label{eq:toepH}
        {\bf \bar H}_{k,u}^{\rm upp} = \left[\begin{array}{cccccc}
        0      &~~\cdots~~  & 0      &\!\!\! h_{k,u}^k[N_{k}^\prime{-}1] &\!\! \cdots &\!\! h_{k,u}^k[L_{\rm I}] \\
        \vdots & 0          & ~      &\!\!\! 0                                &\!\! \ddots &\!\! \vdots \\
        \vdots & ~          & \ddots &\!\!\! ~                                &\!\! \ddots &\!\! h_{k,u}^k[N_{k}^\prime{-}1] \\
        \vdots & ~          & ~      &\!\!\! \ddots                           &\!\! ~      &\!\! 0 \\
        \vdots & ~          & ~      &\!\!\! ~                                &\!\! \ddots &\!\! \vdots \\
        0      & 0          & \cdots &\!\!\! \cdots                           &\!\! \cdots &\!\! 0 \\
        \end{array}\right]~~\text{and}~~
        {\bf \bar H}_{k,u}^{\rm low} = \left[\begin{array}{cccccc}
        h_{k,u}^k[N]                \!\!\!\!\!\!&\!\!\!\!\!\! 0             & \cdots   &\!\!\! \cdots                     \!\!\!& \cdots & 0 \\
        \vdots                      \!\!\!\!\!\!&\!\!\!\!\!\! h_{k,u}^k[N]  & 0        &\!\!\! ~                          \!\!\!& ~      & \vdots \\
        h_{k,u}^k[L_{k,k}{-}1]  \!\!\!\!\!\!&\!\!\!\!\!\! ~             & \ddots   &\!\!\! \ddots                     \!\!\!& ~      & \vdots \\
        0                           \!\!\!\!\!\!&\!\!\!\!\!\! \ddots        & ~        &\!\!\! \ddots                     \!\!\!& \ddots & \vdots \\
        \vdots                      \!\!\!\!\!\!&\!\!\!\!\!\! \ddots        & \ddots   &\!\!\! ~                          \!\!\!& \ddots & 0 \\
        0                           \!\!\!\!\!\!&\!\!\!\!\!\! \cdots        & 0        &\!\!\! h_{k,u}^k[L_{k,k}{-}1] \!\!\!& \cdots & h_{k,u}^k[N] \\
        \end{array}\right].
    \end{align}
    \hrulefill
    \end{figure*}
\fi
Note that when $N\geq L_{k,k}$, ${\bf \bar H}_{k,u}^{\rm low} = {\bf 0}_{(L_{\rm I}-1) \times (L_{\rm I}-1)}$ by the definition of \eqref{eq:toepH}.

Plugging \eqref{eq:desiredH} into \eqref{eq:cp_remove}, we have
\begin{align}\label{eq:cp_remove2}
    {\bf \bar y}_k^b
    = &\sum_{u=1}^{U_k^\prime} \left({\bf \bar H}_{k,u}^{\rm NC} + {\bf \bar H}_{k,u}^{\rm C} \right){\bf F}_{k} {\bf s}_{k,u}^b
     + \sum_{i=1,i\neq k}^K\sum_{u=1}^{U_i^\prime} {\bf \bar H}_{i,u}^k{\bf F}_{k} {\bf s}_{i,u}^b + {\bf \bar z}_k^b,
%    &+ \sum_{u=1}^{U_k^\prime} {\bf \bar H}_{k,u}^{\rm C}{\bf F}_{k} {\bf s}_{k,u}^b
%        + \sum_{i=1,i\neq k}^K\sum_{u=1}^{U_i^\prime} {\bf \bar H}_{i,u}^k{\bf F}_{k} {\bf s}_{i,u}^b .
\end{align}
for $k\in \mathcal{K}$ and $b\in\{1,2,\ldots, B\}$.
As seen in \eqref{eq:interferenceH} and \eqref{eq:cp_remove2}, the channel matrices of ICI links are circulant matrices, while the channel matrices of desired links are the superposition of circulant and non-circulant matrices.
%    ${\bf \bar H}_{i,u}^{k}$ is a circulant matrix for $i \neq k$, while ${\bf \bar H}_{k,u}^{k}$ consists of both circulant and non-circulant matrices.
This difference is due to the fact that the cyclic-prefix length is selected as $ L_{\rm I}-1$ such that $L_{k,i}-1 \leq L_{\rm I}-1 < L_{k,k}-1$ for $i\neq k$ and $i,k\in\mathcal{K}$. The use of the cyclic prefix with this specific length creates \textit{the structural relativity of the channel matrices between desired links and ICI links.}

%%%%%%%%%%%%%%%%%%%%%%%%%%%%%%%%%%%%%%%%%
\vspace{0.2cm}
\textbf{Inter-cell-interference cancellation:}
We explain a ICI cancellation method that harnesses the relativity in the matrix structure between desired and ICI links. We start by providing a lemma that is essential for our proof.
\vspace{0.1cm}
\begin{lem}\label{Lemma1}
    A circulant matrix ${\bf C}\in\mathbb{C}^{n\times n}$ is decomposed as
    \begin{align}
        {\bf C}= {\bf F}{\bm \Lambda}{\bf F}^H,
    \end{align}
    where ${\bf F}=[{\bf f}_1,{\bf f}_2,\ldots,{\bf f}_n]\in\mathbb{C}^{n\times n}$ is the $n$-point IDFT matrix
    whose $k$-th column vector is defined as ${\bf f}_{k}=\frac{1}{\sqrt {n}}\left[1,\omega^{k-1},\omega^{2(k-1)},\ldots ,\omega^{(n-1)(k-1)}\right]^H$,
%    \begin{align}\label{eq:idftcolumn}
%        {\bf f}_{k}=\frac{1}{\sqrt {n}}\left[1,~\omega^{k-1},~\omega^{2(k-1)},~\ldots ,~\omega^{(n-1)(k-1)}\right]^H,
%    \end{align}
    with $\omega=\exp \left({-j\frac {2\pi}{n}}\right)$ for $k=\{1,2,\ldots ,n\}$.
\end{lem}
\begin{IEEEproof}
    See \cite{Golub:96}.
\end{IEEEproof}
\vspace{0.1cm}
As seen in \eqref{eq:interferenceH} and \eqref{eq:cp_remove2}, the channel matrices of ICI links are circulant. Therefore, from Lemma~\ref{Lemma1}, the columns of the precoding matrix ${\bf F}_{k}$ are the eigenvectors of these ICI channel matrices. This implies that all the ICI signals in \eqref{eq:cp_remove2} are aligned in the subspace formed by ${\bf F}_{k}$, i.e., $\text{span}\left( {\bf \bar H}_{i,u}^k{\bf F}_{k} \right) = \text{span}\left({\bf F}_{k}\right)$ for $i\neq k$ and $i,k\in \mathcal{K}$.
%    \begin{align}\label{eq:span2}
%        \text{span}\left( {\bf \bar H}_{i,u}^k{\bf F}_{k} \right) = \text{span}\left({\bf F}_{k}\right),~~\text{for}~~ i\neq k ~\text{and}~ i,k\in \mathcal{K}.
%    \end{align}
As a result, we can eliminate all the ICI signals by projecting ${\bf \bar y}_k^b$ in \eqref{eq:cp_remove2} onto the orthogonal subspace of ${\bf F}_{k}$.
To this end, we use a receive combining matrix defined as ${\bf W} = \big[{\bf f}_{M_{\rm D}+1},\cdots,{\bf f}_{N} \big]^H \in\mathbb{C}^{(N-M_{\rm D}) \times N}$.
Then the effective received signal vector of the $k$-th BS during the $b$-th subblock transmission, namely ${\bf \tilde y}_k^b \in \mathbb{C}^{N-M_{\rm D}}$, is given by
    \begin{align}\label{eq:effreceived}
        {\bf \tilde y}_k^b
        &= {\bf W}{\bf \bar y}_k^b
            = \sum_{u=1}^{U_k^\prime} {\bf W}{\bf \bar H}_{k,u}^{\rm NC}{\bf F}_{k} {\bf s}_{k,u}^b + {\bf W}{\bf \bar z}_k^b,
%        &= \sum_{u=1}^{U_k^\prime-1} {\bf W}{\bf \bar H}_{k,u}^{\rm NC}{\bf F}_{k} {\bf s}_{k,u}^b + {\bf W}{\bf \bar H}_{k,U_k^\prime}^{\rm NC}\bar{\bf F}_{k} \bar{\bf s}_{k,U_k^\prime}^b + {\bf W}{\bf \bar z}_k^b,
%        &= {\bf F_{\mathcal{S}^{\rm c}}}\sum_{u=1}^{U_k^\prime} \sum_{m=1}^{M_{k,u}} {\bf \bar H}_{k,u}^{\rm NC}{\bf f}_{m} {s}_{k,u}^b(m) + {\bf \tilde z}_k^b,
    \end{align}
    for $k\in \mathcal{K}$ and $b\in\{1,2,\ldots, B\}$,
Now, the effective received signal vector in \eqref{eq:effreceived} only contains the transmitted signals from the associating users without ICI.

%%%%%%%%%%%%%%%%%%%%%%%%%%%%%%%%%%%%%%%%%
\vspace{0.2cm}
\textbf{Decodability of subblock data:}
%To accomplish our proof, we need to show the decodability of the data symbols in each subblock transmission.
To accomplish our proof, we show the decodability of the data symbols in each subblock transmission.
First, we find the equivalent representation of \eqref{eq:effreceived} in which an effective channel is represented as a MIMO channel;
    then show that this effective channel matrix has full rank.

To simplify the expression in \eqref{eq:effreceived}, we define an effective channel matrix ${\tilde{\bf H}_k} \in \mathbb{C}^{(N-M_{\rm D})\times U_k^\prime M_k}$ as follows:
    \begin{align}\label{eq:effchannel}
        \tilde{\bf H}_k
        = \left[\begin{array}{c}
            {\bf W}{\bf \bar H}_{k,1}^{\rm NC}{\bf F}_{k},~{\bf W}{\bf \bar H}_{k,2}^{\rm NC}{\bf F}_{k},~\cdots,~{\bf W}{\bf \bar H}_{k,U_k^\prime}^{\rm NC}{\bf F}_k \\
        \end{array}\right].
    \end{align}
    %where ${\bf F}_k^\prime = [{\bf f}_1,{\bf f}_2,\ldots,{\bf f}_{M_k^\prime}]\in\mathbb{C}^{N \times M_k^\prime}$.
Then the effective received signal vector in \eqref{eq:effreceived} simplifies to
\begin{align}\label{eq:effreceived2}
    {\bf \tilde y}_k^b = \tilde{\bf H}_k{\bf s}_k^b +  \tilde{\bf  z}_k^b,
\end{align}
where
    ${\bf s}_k^b = \big[{\bf s}_{k,1}^b,{\bf s}_{k,2}^b, \cdots,{\bf s}_{k,U_k^\prime}^b \big]^{\top}\in \mathbb{C}^{U_k^\prime M_k}$
    is the total data symbol vector received at the $k$-th BS during the $b$-th subblock transmission, and
    ${\bf \tilde z}_k^b={\bf W}{\bf \bar z}_k^b \in \mathbb{C}^{N-M_{\rm D}}$ is an effective noise vector.
Note that the distribution of ${\bf \tilde z}_k^b$ is invariant with ${\bf \bar z}_k^b$ because ${\bf W}$ is a unitary transformation matrix.

Since the expression in \eqref{eq:effreceived2} is equivalent to a simple MIMO system,
    to guarantee the decodability of ${\bf s}_k^b$ in \eqref{eq:effreceived2},
    we only need to show whether the rank of $\tilde{\bf H}_k$ equals $U_k^\prime M_k$ which is the number of data symbols sent by the users in cell $k$. The following lemma essentially shows our decodability result.
\vspace{0.2cm}
\begin{lem}\label{Lemma2}
    The rank of $\tilde{\bf H}_k$ defined in \eqref{eq:effchannel} is
        \begin{align}\label{eq:lem2:cond}
            {\rm rank}\left(\tilde{\bf H}_k\right) = U_k^\prime M_k,~~\text{for}~~k \in \mathcal{K}.
        \end{align}
\end{lem}
\begin{IEEEproof}
    See Appendix A.
\end{IEEEproof}
\vspace{0.2cm}
\noindent Lemma~\ref{Lemma2} implies that for sufficiently large SNR, all $U_k^\prime M_k$ data symbols in ${\bf s}_k^b$ can be reliably decoded.
For example, we can apply MLD or ZFD methods to detect ${\bf s}_k^b$ from \eqref{eq:effreceived2}.
%Then, for sufficiently large SNR, all $U_k^\prime$ data symbols in ${\bf s}_k^b$ can be reliably decoded by applying a ZF detection technique to \eqref{eq:effreceived2}. Consequently, the $k$-th BS achieves DoF of $d_{k}=\frac{B(L_{{\rm D},k} - L_{\rm I})}{B(N +L_{\rm I}-1)+ L_{\rm D}-1}$.

%%%%%%%%%%%%%%%%%%%%%%%%%%%%%%%%%%%%%%%%%
\vspace{0.2cm}
\textbf{Inter-subblock-interference cancellation:}
We have shown that $U_k^\prime M_k$ data symbols are decodable for each subblock transmission, by assuming that there is no inter-subblock interference (ISBI). Unfortunately, ISBI between two subsequent subblocks is unpreventable because the length of the cyclic prefix is shorter than the number of CIR taps for desired links. Thus, a cancellation method of ISBI is needed for multiple subblock transmissions.

 After discarding $\max\{L_{\rm D},L_{\rm I}\}-1$ zeros at the end of the transmission block, we concatenate the received vectors from all subblock transmissions. Then, when ignoring noise, the total input-output relationship during an entire block transmission is
\begin{align}\label{eq:totalblock}
    \left[ \begin{array}{c}
    {\bf \tilde y}_k^1 \\ {\bf \tilde y}_k^2 \\ \vdots \\ {\bf \tilde y}_k^{B}  \\
    \end{array}\right]
    &= \left[\begin{array}{ccccc}
             \tilde{\bf H}_k           & {\bf 0}^{\rm sub}     & \cdots    & \cdots                    & {\bf 0}^{\rm sub} \\
             %& & & & \\
             \tilde{\bf H}_k^{\rm sub} & \tilde{\bf H}_k           & \ddots    & \ddots                        & \vdots \\
             %& & & & \\
             {\bf 0}^{\rm sub}        & \tilde{\bf H}_k^{\rm sub} & \ddots    & \ddots                        & \vdots \\
             %& & & & \\
             \vdots                   & \ddots                             & \ddots    & \tilde{\bf H}_k           & {\bf 0}^{\rm sub}  \\
             %& & & & \\
             {\bf 0}^{\rm sub}        &   \cdots                      & {\bf 0}^{\rm sub}   & \tilde{\bf H}_k^{\rm sub} & \tilde{\bf H}_k \\
        \end{array}\right]
        \left[\begin{array}{c}
            {\bf s}_k^1 \\ {\bf s}_k^2 \\  \vdots \\ {\bf s}_k^{B} \\
        \end{array}\right],
\end{align}
where $\tilde{\bf H}_{k}^{\rm sub}\in \mathbb{C}^{(N-M_{\rm D})\times U_k^\prime M_k}$ is the effective channel matrix for ISBI at the $k$-th BS, and ${\bf 0}^{\rm sub}= {\bf 0}_{(N-M_{\rm D})\times U_k^\prime M_k}$.
By the definition, one can easily verify that $\tilde{\bf H}_{k}^{\rm sub}$ is given by
\begin{align}\label{eq:IBIchannel}
    \tilde{\bf H}_k^{\rm sub} = \left[\begin{array}{c}
            {\bf W}{\bf \bar H}_{k,1}^{\rm sub}{\bf F}_{k},{\bf W}{\bf \bar H}_{k,2}^{\rm sub}{\bf F}_{k}, \cdots, {\bf W}{\bf \bar H}_{k,U_k^\prime}^{\rm sub}{\bf F}_{k} \\
        \end{array}\right].
\end{align}
where ${\bf \bar H}_{k,u}^{\rm sub} \in \mathbb{C}^{N\times N}$ is
\begin{align}
    {\bf \bar H}_{k,u}^{\rm sub} = \left[\begin{array}{cc}
        {\bf 0}_{(N-L_{\rm I}+1)  \times (L_{\rm I}-1)}  &  {\bf \bar H}_{k,u}^{\rm upp}\Big|_{N_{k}^\prime=L_{k,k}} \\
         {\bf 0}_{(L_{\rm I}-1)  \times (L_{\rm I}-1)} &  {\bf 0}_{(L_{\rm I}-1) \times (N-L_{\rm I}+1)} \\
    \end{array}\right].
\end{align}
At the first subblock transmission, there is no ISBI, i.e., ${\bf \tilde y}_k^1= \tilde{\bf H}_k{\bf s}_k^1+{\bf \tilde z}_k^1$,
    so the symbol vector ${\bf s}_k^1$ is reliably decodable for a sufficiently large SNR value.
% Let ${\bf s}_k^b$ be the decoded information vector during the $b$-th subblock transmission.
At the $b$-th subblock transmission for $b\geq 2$, under the premise that ${\bf s}_k^{b-1}$ is reliably decodable,
    it is possible to decode ${\bf s}_k^b$ by subtracting the effect of ${\bf s}_k^{b-1}$ from the received signal ${\bf \tilde y}_k^b$ as follows:
\begin{align}\label{eq:subtract_block}
    {\bf \tilde y}_k^b - \tilde{\bf H}_k^{\rm sub}{\bf s}_k^{b-1} = \tilde{\bf H}_k{\bf s}_k^b +{\bf \tilde z}_k^b.
\end{align}
One practical concern with this successive interference cancellation method is that, when SNR is low, it may suffer from error propagation. Nevertheless, this approach is sufficient to show the DoF result in the high SNR regime.

%%%%%%%%%%%%%%%%%%%%%%%%%%%%%%%%%%%%%%%%%
\vspace{0.2cm}
\textbf{Achievable sum-DoF calculation:}
Applying the above ISBI cancellation strategy over $B$ subblocks recursively,
    the $k$-th BS is capable of decoding $B$ data symbol vectors ${\bf s}_k^1, {\bf s}_k^2, \ldots, {\bf s}_k^{B}$,
    with $T = B{\bar N}+L_{\rm D}-1$ time slots.
For sufficiently large coherence time, $B$ can be taken to be infinity,
    so the achievable DoF of the $k$-th BS is
\begin{align}
    d_k = \lim_{B\rightarrow \infty}\frac{BU_k^\prime M_k}{B(N+L_{\rm I}-1)+\max\{L_{\rm D},L_{\rm I}\}-1} =\frac{U_k^\prime M_k}{N+L_{\rm I}-1}. \label{eq:final}
\end{align}
By plugging $N =\max\left\{L_{\rm D}-L_{\rm I}+M_{\rm D},L_{\rm I}\right\}$ to \eqref{eq:final}, we arrive at the expression in Theorem~\ref{Theorem1}.
\end{IEEEproof}
\vspace{0.3cm}

%%%%%%%%%%%%% Corollary 1
Theorem~\ref{Theorem1} shows the achievable sum-DoF for a general ISI condition, yet it is unwieldy to provide a clear intuition in the result. Considering a symmetric ISI scenario, we simplify Theorem~\ref{Theorem1} to the following Corollary:
\begin{cor} ({\bf Symmetric ISI condition})\label{Corollary1}
    Consider a $K$-cell MAC with symmetric ISI, i.e., $L_{k,k} = L_{{\rm D}}$ and $L_{k,i}= L_{\rm I}$ for $i\neq k$ and $i, k \in\mathcal{K}$.
    When the number of users per cell is larger than $L_{{\rm D}}-L_{\rm I}$ with $L_{\rm D}\geq 2L_{\rm I}$, the achievable sum-DoF of the considered channel is
    \begin{align}\label{eq:cor1:sumDoF}
        d^{\rm MAC}_{\Sigma} =  \left(1-\frac{L_{\rm I}}{L_{\rm D}} \right)K  \rightarrow K,   ~~\text{as}~~\frac{L_{\rm D}}{L_{\rm I}} \rightarrow \infty.
    \end{align}
\end{cor}
\vspace{0.2cm}
\begin{IEEEproof}
    Suppose that $L_{k,k} = L_{{\rm D}}$ and $L_{k,i}= L_{\rm I}$ for $i\neq k$ and $i, k \in\mathcal{K}$.
    If $U_k \geq L_{{\rm D}}-L_{\rm I}$ for $k\in\mathcal{K}$ with $L_{\rm D}\geq 2L_{\rm I}$, from \eqref{eq:thm1:sumDoF}, the achievable sum-DoF is obtained as in \eqref{eq:cor1:sumDoF}.
%    \eqref{eq:thm1:sumDoF}:
%    \begin{align}\label{eq:cor1:sumDoF2}
%        d^{\rm MAC}_{\Sigma} =  \left(1-\frac{L_{\rm I}}{L_{\rm D}} \right)K.
%    \end{align}
%    Consequently, by taking $\frac{L_{\rm D}}{L_{\rm I}} \rightarrow \infty$, we complete the proof.
\end{IEEEproof}
\vspace{0.2cm}

Corollary~\ref{Corollary1} implies that \emph{interference-free DoF per cell} is asymptotically achievable even without CSIT,
    as the ratio of the maximum CIR length of desired links to that of ICI links approaches infinity with a sufficiently large number of users per cell.
    %In other words, a total of $K$ DoF is asymptotically achievable if there exists a sufficient number of users whose channel condition satisfies that $L_{{\rm D}}\gg L_{\rm I}$.
Particularly when the ICI links have the small delay spread, the condition required to achieve $K$ sum-DoF can be further relaxed.
For example, if all ICI links are line-of-sight channels, i.e., $L_{\rm I}=1$,
    the equation in \eqref{eq:cor1:sumDoF} becomes
    \begin{align}
        d^{\rm MAC}_{\Sigma} = \frac{L_{\rm D}-1} {L_{\rm D}}K.
    \end{align}
In this case, nearly $K$ sum-DoF is achievable even when both the number of users and the maximum CIR length of desired links are not so large.

%%%%%%%%%%%%% Corollary 2
With the proposed interference management method, it is also possible to characterize the  achievable sum-spectral efficiency in a closed form, which is given in the following Corollary:
\begin{cor} ({\bf Achievable sum-spectral efficiency})\label{Corollary2}
    Consider a $K$-cell MAC with ISI, each cell with $U_k^\prime$ active uplink users that transmit $M_k$ data symbols respectively. Let $\tilde{\bf H}_k = {\bf Q}_{k}{\bf R}_{k}$ be the QR decomposition of the effective channel matrix $\tilde{\bf H}_k$ in \eqref{eq:effchannel}, where ${\bf Q}_{k}\in\mathbb{C}^{(N-M_{\rm D}) \times U_k^\prime M_k }$ is a semi-unitary matrix such that ${\bf Q}_{k}^H{\bf Q}_{k} = {\bf I}_{U_k^\prime M_k}$,
    and ${\bf R}_{k}\in\mathbb{C}^{U_k^\prime M_k \times U_k^\prime  M_k}$ is an upper-triangular matrix with positive diagonal elements. Then the achievable sum-spectral efficiency of this channel without instantaneous CSIT is
    \begin{align}\label{eq:cor2:rate}
        \sum_{k=1}^{K}\sum_{m=1}^{U_k^\prime M_k} \frac{  \log_2\left(1+\frac{N}{M_k}|r_{k,m}|^2\rho\right)}{N+L_{\rm I}-1},
    \end{align}
    where $r_{k,m}$ is the $m$-th diagonal element of ${\bf R}_{k}$ and ${\rho}=\frac{P}{\sigma^2}$ is SNR.
\end{cor}
\vspace{0.2cm}
\begin{IEEEproof}
	Applying the QR decomposition \cite{Golub:96}, the effective channel matrix $\tilde{\bf H}_k$ in \eqref{eq:subtract_block} can be decomposed as $\tilde{\bf H}_k = {\bf Q}_{k}{\bf R}_{k}$,
where ${\bf Q}_{k}\in\mathbb{C}^{(N-M_{\rm D}) \times U_k^\prime M_k }$ is a semi-unitary matrix such that ${\bf Q}_{k}^H{\bf Q}_{k} = {\bf I}_{U_k^\prime M_k}$,
    and ${\bf R}_{k}\in\mathbb{C}^{U_k^\prime M_k \times U_k^\prime  M_k}$ is an upper-triangular matrix with positive diagonal elements.
By plugging the QR decomposition of $\tilde{\bf H}_k$ to \eqref{eq:subtract_block}, the effective received signal of the $k$-th BS during the $b$-th subblock transmission is rewritten as
    \begin{align}\label{eq:subtract_block2}
        {\bf \tilde y}_k^b - \tilde{\bf H}_k^{\rm sub}{\bf s}_k^{b-1} &= \tilde{\bf H}_k{\bf s}_k^b + {\bf \tilde z}_k^b
        ={\bf Q}_{k}{\bf R}_{k} {\bf s}_{k}^b + {\bf \tilde z}_k^b.
    \end{align}
Under the premise that each user uses the Gaussian signaling, the achievable rate at the $k$-th BS sent over $T=B(N+L_{\rm I}-1)+L_{\rm D}-1$ time slots by zero-forcing successive-interference cancellation is computed as
\begin{align}
    R_k= \sum_{m=1}^{U_k^\prime M_k}\frac{ B  \log_2\left(1+\tilde{\rho}_k|r_{k,m}|^2\right)}{B(N+L_{\rm I}-1)+L_{\rm D}-1},
\end{align}
where $r_{k,m}$ is the $m$-th diagonal element of ${\bf R}_{k}$ and $\tilde{\rho}_k$ is an effective SNR in \eqref{eq:subtract_block2}.
With the power constraint of $\mathbb{E}[|x_{k,u}[n]|^2]=P$, the effective SNR is given by $\tilde{\rho}_k=\frac{NP}{ M_k\sigma^2}$ because $\mathbb{E}[\|{\bf F}_{k}{\bf s}_{k,u}^{b} \|^2]=NP$ for $u \in \mathcal{U}_k$.
Then the achievable sum-spectral efficiency is given by
\begin{align}
    \lim_{B\rightarrow \infty}\sum_{k=1}^KR_k = \sum_{k=1}^{K}\sum_{m=1}^{U_k^\prime M_k} \frac{  \log_2\left(1+\tilde{\rho}_k|r_{k,m}|^2\right)}{N+L_{\rm I}-1}, \label{eq:rates_with_no_CSIT}
\end{align}
as $B$ tends to infinity; this completes the proof.
\end{IEEEproof}
\vspace{0.2cm}

\begin{figure}
\centering \vspace{-0.3cm}
\includegraphics[width=3.1in]{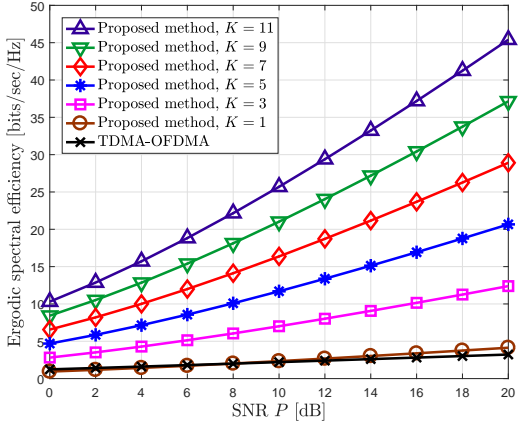} \vspace{-0.3cm}\caption{Comparison of sum-spectral efficiency between TDMA-OFDMA and the proposed interference management method for different $K$. We set $B=10$, $L_{k,k}=L_{\rm D}=8$, $L_{k,i}=L_{\rm I}=2$, and $U_{k}=U=3$ for $i\neq k$ and $i,k\in \mathcal{K}$. Each CIR tap, $h_{i,u}^k[\ell]$, is drawn from $\mathcal{CN}(0,1)$ for all $\ell \in \{1,\ldots,L_{k,i}\}$, $u \in \{1,\ldots,U_i^\prime\}$, and $i,k \in \mathcal{K}$.} \label{fig:SE}
\end{figure}

Corollary~\ref{Corollary2} is useful to gauge the performance benefit of the proposed method in a low SNR regime. As a numerical example, Fig. \ref{fig:SE} shows that the ergodic sum-spectral efficiency of the proposed interference management method increases linearly with $K$, so the proposed method outperforms TDMA-OFDMA in all SNR regimes when $K > 1$.

%%%%%%%%%%%%%%%%%%%%%%
\vspace{1mm}
{\bf Remark 2 (Sum-DoF comparison with the existing work in \cite{Lee:16}):}
The concept of the blind interference management with matrix structuring has originally been proposed in the context of a $K$-cell interference channel with ISI \cite{Lee:16}, where the achievable sum-DoF with the symmetric ISI condition has shown to be
    \begin{align}\label{eq:dof_IC}
        d^{\rm IC}_{\Sigma} = K \frac{L_{{\rm D}} - L_{\rm I}} {\max\left\{2L_{\rm D}-L_{\rm I}-1, 2L_{\rm I}-1 \right\}}\rightarrow \frac{K}{2},
            ~\text{as}~ \frac{L_{\rm D}}{L_{\rm I}} \rightarrow \infty.
    \end{align}
From \eqref{eq:dof_IC}, we can easily see that the achievable sum-DoF for interfering MAC is twice higher than that attained in the interference channel, in an asymptotic sense.
%This additional gain in sum-DoF is obtained from the \emph{multi-user diversity} in MAC, where independent channels are realized from different users.

\ifonecol
\begin{figure}[t]
    \centering \vspace{-3mm}
    \subfigure[$L_{\rm D} > L_{\rm I}$ without propagation delay for ICI link]
    {\epsfig{file=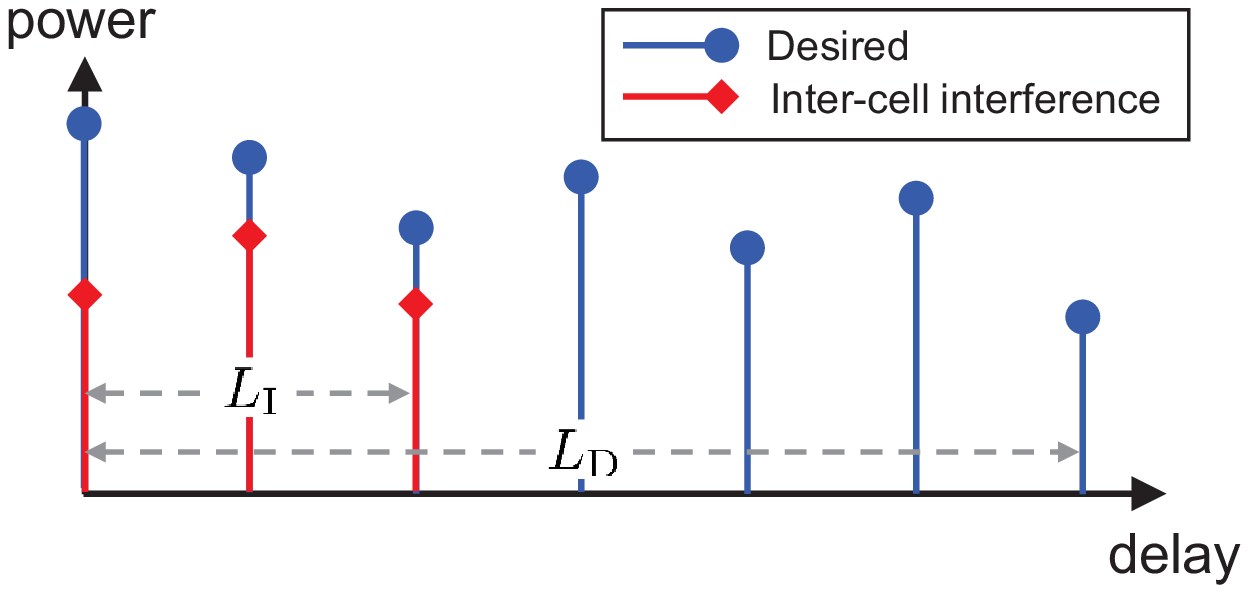, width=5.2cm}} \vspace{-1mm}
    \subfigure[$L_{\rm D} > L_{\rm I}$ with propagation delay for ICI link]
    {\epsfig{file=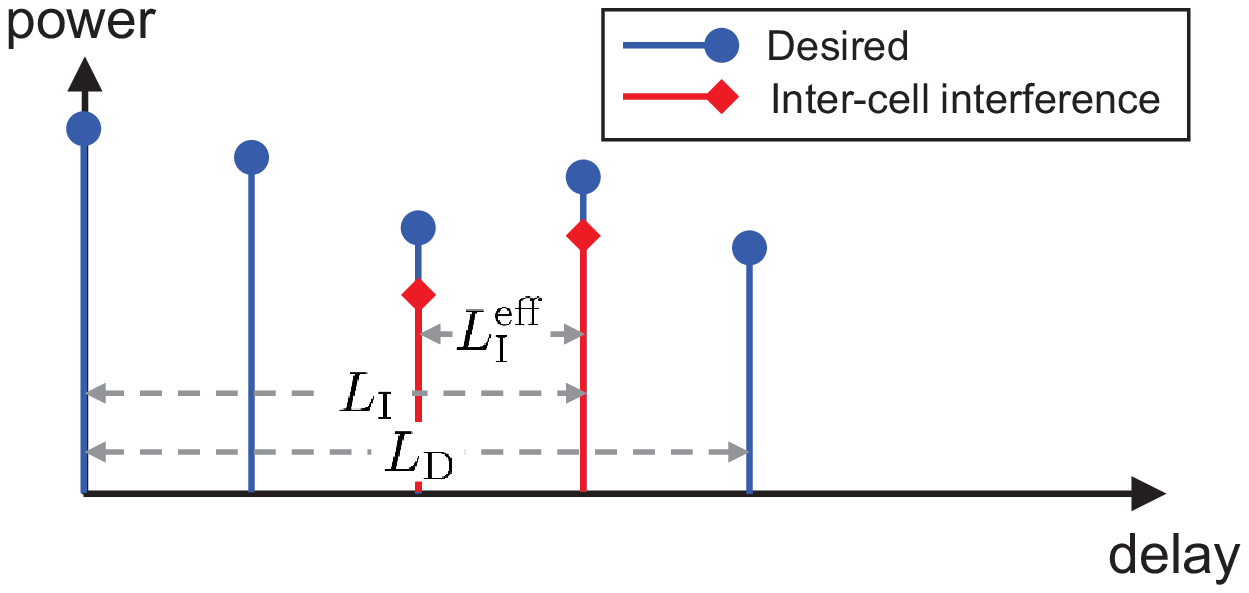, width=5.2cm}} \vspace{-1mm}
    \subfigure[$L_{\rm D} \leq L_{\rm I}$ with propagation delay for ICI link]
    {\epsfig{file=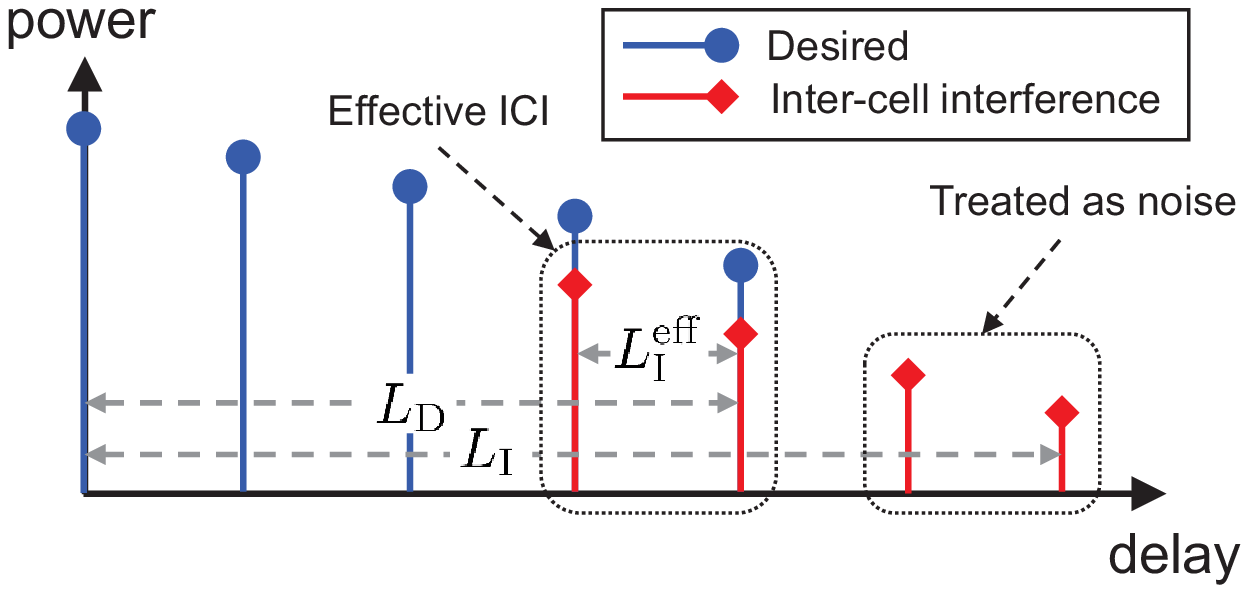, width=5.2cm}} \vspace{-1mm}
    \caption{Three typical power-delay profiles of the channels with propagation delay.}
    \label{fig:extension}
\end{figure}
\else
\begin{figure}[t]
    \centering \vspace{-3mm}
    \subfigure[$L_{\rm D} > L_{\rm I}$ without propagation delay for ICI link]
    {\epsfig{file=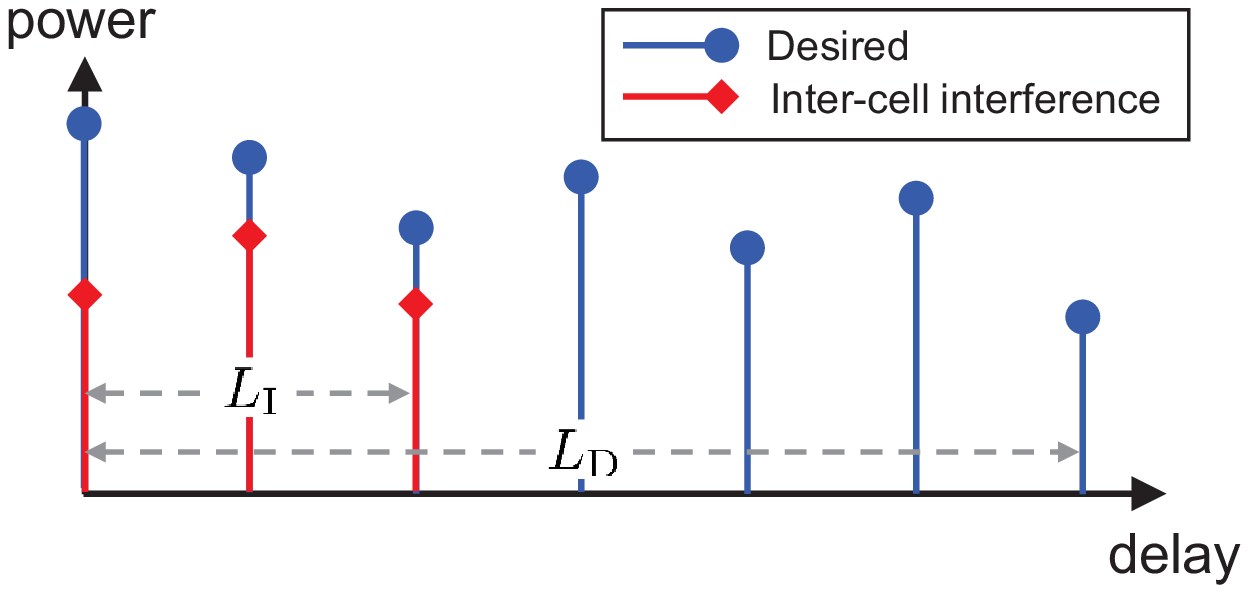, width=6cm}} \vspace{-1mm}
    \subfigure[$L_{\rm D} > L_{\rm I}$ with propagation delay for ICI link]
    {\epsfig{file=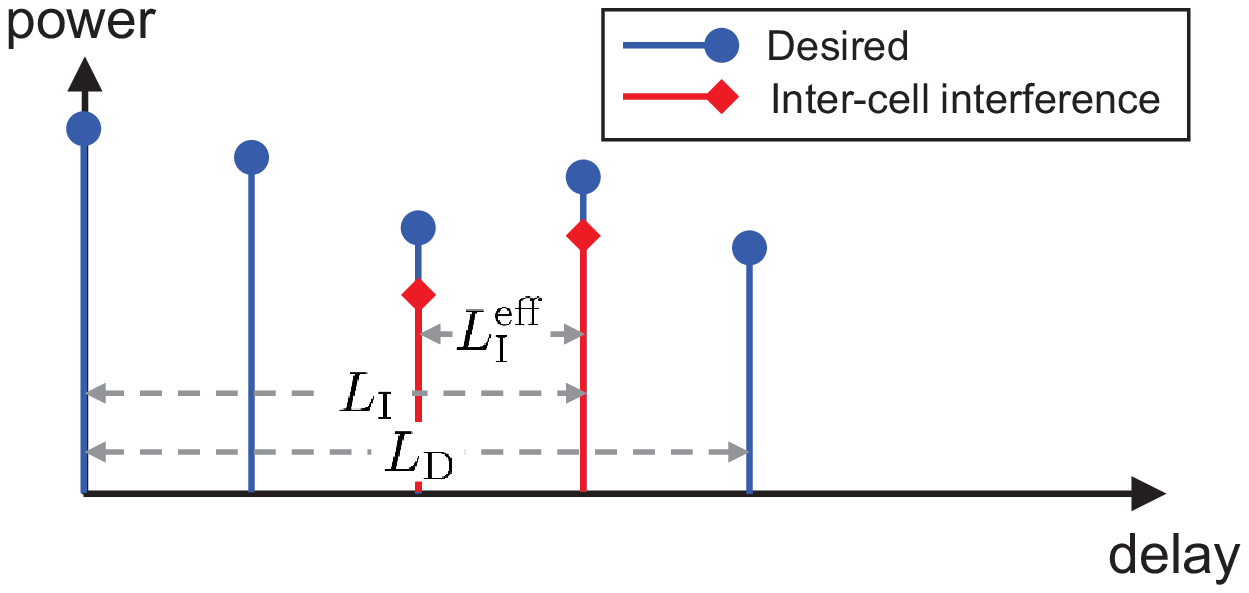, width=6cm}} \vspace{-1mm}
    \subfigure[$L_{\rm D} \leq L_{\rm I}$ with propagation delay for ICI link]
    {\epsfig{file=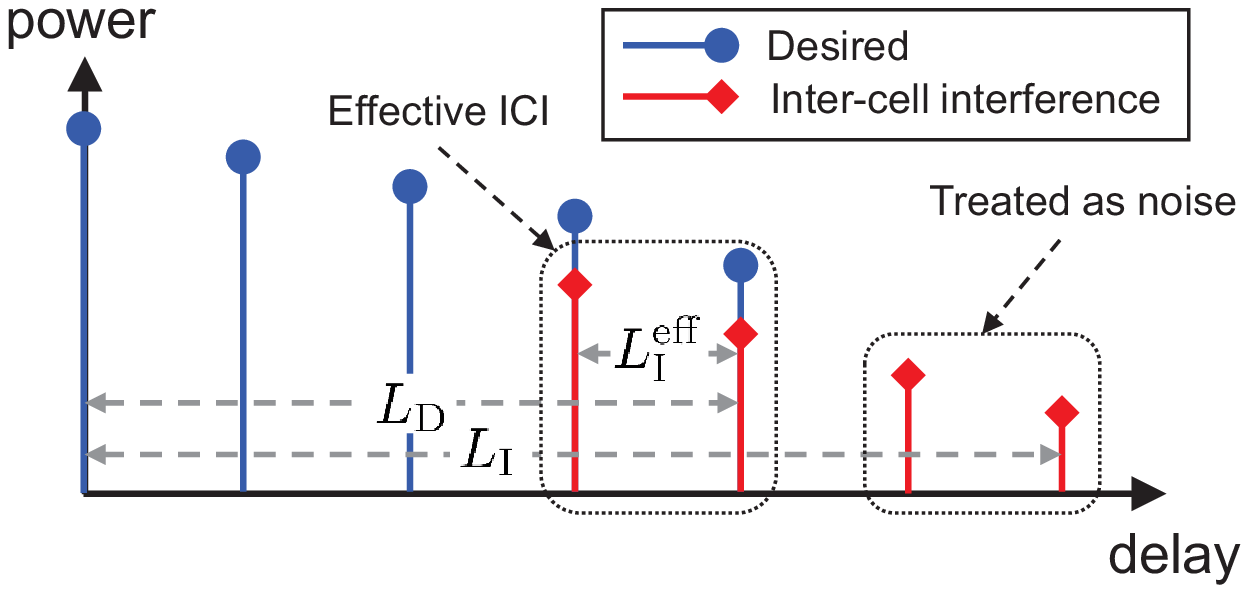, width=6cm}} \vspace{-1mm}
    \caption{Three typical power-delay profiles of the channels with propagation delay.}
    \label{fig:extension}
\end{figure}
\fi

%%%%%%%%%%%%%%%%%%%%%%%%%%%%%%%%%%%%%%%%%%%%%%%%%%%%%%%%%%%%%%%%%%
\section{Discussions}
For simplicity, the proposed interference management method has been presented under the assumption that the delay spread of the desired links is larger than the maximum delay spread of ICI links, i.e., $L_{k,k} > L_{\rm I}$, and all the channel values are non-zeros, i.e., $h_{i,u}^k[\ell]\neq 0$ for $\ell\in\{1,2,\ldots, L_{k,i}\}$, $u\in\mathcal{U}_i$, and $i,k\in\mathcal{K}$. This assumption may not be hold in some wireless environments because it ignores the propagation delay between the desired and interfering signals.

In this section, we discuss how to modify the proposed interference management method for two different cases: 1)
    $L_{\rm D} > L_{\rm I}$ and 2) $L_{\rm D} \leq L_{\rm I}$, considering the relative propagation delay between the desired and ICI links.

\subsection{Effect of Propagation Delay when $L_{\rm D} > L_{\rm I}$}
When propagation delay for ICI link exists as depicted in Fig.~\ref{fig:extension}(b), the sum-DoF can be further improved by exploiting ICI-free signals during the delay. In what follows, we demonstrate this improvement by using an illustrative example.

\vspace{0.1cm}
{\bf Example 2 ($K$-cell MAC with Three Users):} Consider a $K$-cell MAC with three users per cell when $L_{k,k} =L_{\rm D}=5$ for $u\in \mathcal{U}_k$ and $L_{k,i} =L_{\rm I}=4$ for $u \in \mathcal{U}_i$, $i\neq k$, and $i, k\in\mathcal{K}$. Let $L_{\rm I,d}$ be the number of CIR taps during ICI delay offset, then $L_{\rm I}^{\rm eff} = L_{\rm I}- L_{\rm I,d}$ becomes the effective (actual) number of CIR taps for ICI links in a system. With this notation, we consider a delayed-ICI case with $L_{\rm I}^{\rm eff} = 2$, i.e., $h_{i,u}^k[0]=h_{i,u}^k[1]=0$ for $u \in \mathcal{U}_i$, $i\neq k$, and $i, k\in\mathcal{K}$. The considered scenario is depicted in Fig.~\ref{fig:extension}(b). In this scenario, we will show that each BS reliably decodes three data symbols with seven time slots, i.e., $d_{\Sigma}=\frac{3K}{7}$, by using the proposed interference management method with a minor modification.

Using the transmission strategy of the proposed interference management method, suppose that all users use the first column of the $4$-point IDFT matrix to convey one data symbol, i.e., ${\bf \bar x}_{k,u} = {\bf f}_1 s_{k,u} \in \mathbb{C}^{4}$, where ${\bf f}_1 = 0.5[1,1,1,1]^{\top}$. The transmitted signal vector is generated by adding the cyclic prefix with the length of $L_{\rm I}-1 = 3$ to $\bar{\bf x}_{k,u}$. Then the received signal vector at the $k$-th BS during seven time slots is given as in \eqref{eq:ex2:received} (see the top of the page).
\ifonecol
\begin{figure*}
\vspace{-3mm}{\scriptsize\begin{align}\label{eq:ex2:received}
    \left[\!\begin{array}{c}
            {y}_{k}[1] \\ {y}_{k}[2] \\ {y}_{k}[3] \\ {y}_{k}[4] \\ {y}_{k}[5]  \\
            {y}_{k}[6] \\ {y}_{k}[7]
        \end{array}\!\right]
    = & \sum_{u=1}^3 \left[\!\begin{array}{ccccccc}
            {h}_{k,u}^k[0] \!&\! 0 \!&\! 0 \!&\! 0 \!&\! 0 \!&\! 0 \!&\! 0  \\
            {h}_{k,u}^k[1] \!&\! {h}_{k,u}^k[0] \!&\! 0 \!&\! 0 \!&\! 0 \!&\! 0 \!&\! 0  \\
            {h}_{k,u}^k[2] \!&\! {h}_{k,u}^k[1] \!&\! {h}_{k,u}^k[0] \!&\! 0 \!&\! 0 \!&\! 0 \!&\! 0\\
            {h}_{k,u}^k[3] \!&\! {h}_{k,u}^k[2] \!&\! {h}_{k,u}^k[1] \!&\! {h}_{k,u}^k[0] \!&\! 0 \!&\! 0 \!&\! 0 \\
            {h}_{k,u}^k[4] \!&\! {h}_{k,u}^k[3] \!&\! {h}_{k,u}^k[2] \!&\! {h}_{k,u}^k[1] \!&\! {h}_{k,u}^k[0] \!&\! 0 \!&\! 0 \\
            0 \!&\! {h}_{k,u}^k[4] \!&\! {h}_{k,u}^k[3] \!&\! {h}_{k,u}^k[2] \!&\! {h}_{k,u}^k[1] \!&\! {h}_{k,u}^k[0] \!&\! 0 \\
            0 \!&\! 0 \!&\! {h}_{k,u}^k[4] \!&\! {h}_{k,u}^k[3] \!&\! {h}_{k,u}^k[2] \!&\! {h}_{k,u}^k[1] \!&\! {h}_{k,u}^k[0] \\
        \end{array}\right]
        \left[\!\begin{array}{c}
            {x}_{k,u}[2] \\ {x}_{k,u}[3] \\ {x}_{k,u}[4]  \\ {x}_{k,u}[1] \\ {x}_{k,u}[2] \\ {x}_{k,u}[3] \\ {x}_{k,u}[4] \\
        \end{array}\!\right] \nonumber \\
    &+ \sum_{i\neq k}\sum_{u=1}^3 \left[\!\begin{array}{ccccc}
            0 \!&\! 0 \!&\! 0 \!&\! 0 \!&\! 0 \\
            0 \!&\! 0 \!&\! 0 \!&\! 0 \!&\! 0 \\
            {h}_{i,u}^k[2] \!&\! 0 \!&\! 0 \!&\! 0 \!&\! 0 \\
            {h}_{i,u}^k[3] \!&\! {h}_{i,u}^k[2] \!&\! 0 \!&\! 0 \!&\! 0 \\
            0 \!&\! {h}_{i,u}^k[3] \!&\! {h}_{i,u}^k[2] \!&\! 0 \!&\! 0 \\
            0 \!&\! 0 \!&\! {h}_{i,u}^k[3] \!&\! {h}_{i,u}^k[2] \!&\! 0 \\
            0 \!&\! 0 \!&\! 0 \!&\! {h}_{i,u}^k[3] \!&\! {h}_{i,u}^k[2] \\
        \end{array}\!\right]
        \left[\!\begin{array}{c}
            {x}_{i,u}[2] \\ {x}_{i,u}[3] \\ {x}_{i,u}[4]  \\ {x}_{i,u}[1] \\ {x}_{i,u}[2]
        \end{array}\!\right]
        + \left[\!\begin{array}{c}
            {z}_{k}[1] \\ {z}_{k}[2] \\ {z}_{k}[3] \\ {z}_{k}[4] \\ {z}_{k}[5]  \\
            {z}_{k}[6] \\ {z}_{k}[7]
        \end{array}\!\right].
\end{align}}
\hrulefill
\end{figure*}
\else
\begin{figure*}
\vspace{-3mm}{\scriptsize\begin{align}\label{eq:ex2:received}
    \left[\!\begin{array}{c}
            {y}_{k}[1] \\ {y}_{k}[2] \\ {y}_{k}[3] \\ {y}_{k}[4] \\ {y}_{k}[5]  \\
            {y}_{k}[6] \\ {y}_{k}[7]
        \end{array}\!\right]
    = & \sum_{u=1}^3 \left[\!\begin{array}{ccccccc}
            {h}_{k,u}^k[0] \!&\! 0 \!&\! 0 \!&\! 0 \!&\! 0 \!&\! 0 \!&\! 0  \\
            {h}_{k,u}^k[1] \!&\! {h}_{k,u}^k[0] \!&\! 0 \!&\! 0 \!&\! 0 \!&\! 0 \!&\! 0  \\
            {h}_{k,u}^k[2] \!&\! {h}_{k,u}^k[1] \!&\! {h}_{k,u}^k[0] \!&\! 0 \!&\! 0 \!&\! 0 \!&\! 0\\
            {h}_{k,u}^k[3] \!&\! {h}_{k,u}^k[2] \!&\! {h}_{k,u}^k[1] \!&\! {h}_{k,u}^k[0] \!&\! 0 \!&\! 0 \!&\! 0 \\
            {h}_{k,u}^k[4] \!&\! {h}_{k,u}^k[3] \!&\! {h}_{k,u}^k[2] \!&\! {h}_{k,u}^k[1] \!&\! {h}_{k,u}^k[0] \!&\! 0 \!&\! 0 \\
            0 \!&\! {h}_{k,u}^k[4] \!&\! {h}_{k,u}^k[3] \!&\! {h}_{k,u}^k[2] \!&\! {h}_{k,u}^k[1] \!&\! {h}_{k,u}^k[0] \!&\! 0 \\
            0 \!&\! 0 \!&\! {h}_{k,u}^k[4] \!&\! {h}_{k,u}^k[3] \!&\! {h}_{k,u}^k[2] \!&\! {h}_{k,u}^k[1] \!&\! {h}_{k,u}^k[0] \\
        \end{array}\right]
        \left[\!\begin{array}{c}
            {x}_{k,u}[2] \\ {x}_{k,u}[3] \\ {x}_{k,u}[4]  \\ {x}_{k,u}[1] \\ {x}_{k,u}[2] \\ {x}_{k,u}[3] \\ {x}_{k,u}[4] \\
        \end{array}\!\right]
    + \sum_{i\neq k}\sum_{u=1}^3 \left[\!\begin{array}{ccccc}
            0 \!&\! 0 \!&\! 0 \!&\! 0 \!&\! 0 \\
            0 \!&\! 0 \!&\! 0 \!&\! 0 \!&\! 0 \\
            {h}_{i,u}^k[2] \!&\! 0 \!&\! 0 \!&\! 0 \!&\! 0 \\
            {h}_{i,u}^k[3] \!&\! {h}_{i,u}^k[2] \!&\! 0 \!&\! 0 \!&\! 0 \\
            0 \!&\! {h}_{i,u}^k[3] \!&\! {h}_{i,u}^k[2] \!&\! 0 \!&\! 0 \\
            0 \!&\! 0 \!&\! {h}_{i,u}^k[3] \!&\! {h}_{i,u}^k[2] \!&\! 0 \\
            0 \!&\! 0 \!&\! 0 \!&\! {h}_{i,u}^k[3] \!&\! {h}_{i,u}^k[2] \\
        \end{array}\!\right]
        \left[\!\begin{array}{c}
            {x}_{i,u}[2] \\ {x}_{i,u}[3] \\ {x}_{i,u}[4]  \\ {x}_{i,u}[1] \\ {x}_{i,u}[2]
        \end{array}\!\right]
        + \left[\!\begin{array}{c}
            {z}_{k}[1] \\ {z}_{k}[2] \\ {z}_{k}[3] \\ {z}_{k}[4] \\ {z}_{k}[5]  \\
            {z}_{k}[6] \\ {z}_{k}[7]
        \end{array}\!\right].
\end{align}}
\hrulefill
\end{figure*}
\fi
It can be seen in \eqref{eq:ex2:received} that the received signal vector is divided into two types of signals: \emph{ICI-free} and \emph{ICI-corrupted} signals. In this example, the first two received signals ($y_k[1]$ and $y_k[2]$) correspond to ICI-free signals, while the remaining five received signals correspond to ICI-corrupted signals.  Our goal is to harness these ICI-free signals to increase the rank of the effective channel matrix ${\bf \tilde H}_{k}$. This can be achieved by multiplying a first-stage receive combining matrix defined as
\begin{align}
    {\bf W}_{\rm 1} = \left[\!\begin{array}{ccccccc}
            0 & 0 & 0 & 1 & 0 & 0 & 0 \\
            1 & 0 & 0 & 0 & 1 & 0 & 0 \\
            0 & 1 & 0 & 0 & 0 & 1 & 0 \\
            0 & 0 & 0 & 0 & 0 & 0 & 1 \\
        \end{array}\!\right],
\end{align}
to the received signal vector. Note that this multiplication also substitutes the process for discarding the cyclic prefix. The received signal after multiplying ${\bf W}_{1}$ is represented in \eqref{eq:ex2:received2} (see the top of the page). As seen in \eqref{eq:ex2:received2}, ${\bf \bar H}_{k,u}^{\rm C}$ and ${\bf \bar H}_{i,u}^k$ for $i\neq k$ are circulant matrices,
    while ${\bf \bar H}_{k,u}^{\rm NC}$ is a non-circulant one.
\ifonecol
\begin{figure*}
\vspace{-3mm}{\scriptsize\begin{align}\label{eq:ex2:received2}
    {\bf W}_{1}{\bf y}_k = \underbrace{\left[\!\begin{array}{c}
            y_k[4] \\ y_k[1]{+}y_k[5] \\ y_k[2]{+}y_k [6] \\ y_k[7] \\
        \end{array}\!\right]}_{{\bf \bar y}_k}
    = & \sum_{u=1}^3 \Vast\{\underbrace{\left[\!\begin{array}{cccc}
                -{h}_{k,u}^k[4] \!&\!\! 0              \!&\!\! 0          \!&\! 0 \\
            0               \!&\!\! {h}_{k,u}^k[0] \!&\!\! 0              \!&\! 0 \\
            0               \!&\!\! {h}_{k,u}^k[1] \!&\!\! {h}_{k,u}^k[0] \!&\! 0 \\
            0               \!&\!\! 0              \!&\!\! 0              \!&\! 0 \\
        \end{array}\right]}_{{\bf \bar H}_{k,u}^{\rm NC}}
        +  \underbrace{\left[\!\begin{array}{cccc}
            {h}_{k,u}^k[0]{+}{h}_{k,u}^k[4] \!\!&\!\!\!\!\!\! {h}_{k,u}^k[3] \!&\!\!\!\!\! {h}_{k,u}^k[2] \!&\!\!\!\!\! {h}_{k,u}^k[1]  \\
            {h}_{k,u}^k[1] \!\!&\!\! \!\!\!\!{h}_{k,u}^k[0]{+}{h}_{k,u}^k[4] \!&\!\!\!\!\! {h}_{k,u}^k[3] \!&\!\!\!\!\! {h}_{k,u}^k[2] \\
            {h}_{k,u}^k[2] \!\!&\!\!\!\!\!\! {h}_{k,u}^k[1] \!&\!\!\!\!\! {h}_{k,u}^k[0]{+}{h}_{k,u}^k[4] \!&\!\!\!\!\!  {h}_{k,u}^k[3] \\
            {h}_{k,u}^k[3] \!\!&\!\!\!\!  \!\!{h}_{k,u}^k[2] \!&\!\!\!\!\! {h}_{k,u}^k[1] \!&\!\!\!\!\! {h}_{k,u}^k[0]{+}{h}_{k,u}^k[4]\\
        \end{array}\!\right]}_{{\bf \bar H}_{k,u}^{\rm C}} \Vast\} {\bf f}_1s_{k,u} \nonumber \\
    &~~~+ \sum_{i\neq k}\sum_{u=1}^3 \underbrace{\left[\!\begin{array}{ccccc}
            0              \!&\! {h}_{i,u}^k[3] \!&\! {h}_{i,u}^k[2] \!&\! 0 \\
            0              \!&\! 0              \!&\! {h}_{i,u}^k[3] \!&\! {h}_{i,u}^k[2] \\
            {h}_{i,u}^k[2] \!&\! 0              \!&\! 0              \!&\! {h}_{i,u}^k[3] \\
            {h}_{i,u}^k[3] \!&\! {h}_{i,u}^k[2] \!&\! 0              \!&\! 0 \\
        \end{array}\!\right]}_{{\bf \bar H}_{i,u}^k}{\bf f}_1s_{i,u}
        + \underbrace{\left[\!\begin{array}{c}
            {z}_{k}[4] \\ {z}_{k}[1]{+}{z}_{k}[5] \\ {z}_{k}[2]{+}z_k [6] \\ {z}_{k}[7] \\
        \end{array}\!\right]}_{{\bf \bar z}_k}.
\end{align}}
\hrulefill
\end{figure*}
\else
\begin{figure*}
\vspace{-3mm}{\scriptsize\begin{align}\label{eq:ex2:received2}
    {\bf W}_{1}{\bf y}_k = \left[\!\begin{array}{c}
            y_k[4] \\ y_k[1]{+}y_k[5] \\ y_k[2]{+}y_k [6] \\ y_k[7] \\
        \end{array}\!\right]
    = & \sum_{u=1}^3 \underbrace{\left[\!\begin{array}{cccc}
                -{h}_{k,u}^k[4] \!&\! 0              \!&\! 0              \!&\! 0 \\
            0               \!&\! {h}_{k,u}^k[0] \!&\! 0              \!&\! 0 \\
            0               \!&\! {h}_{k,u}^k[1] \!&\! {h}_{k,u}^k[0] \!&\! 0 \\
            0               \!&\! 0              \!&\! 0              \!&\! 0 \\
        \end{array}\right]}_{{\bf \bar H}_{k,u}^{\rm NC}} {\bf f}_1s_{k,u}
        + \sum_{u=1}^3 \underbrace{\left[\!\begin{array}{cccc}
            {h}_{k,u}^k[0]{+}{h}_{k,u}^k[4] \!\!&\!\! {h}_{k,u}^k[3] \!&\! {h}_{k,u}^k[2] \!&\! {h}_{k,u}^k[1]  \\
            {h}_{k,u}^k[1] \!\!&\!\! {h}_{k,u}^k[0]{+}{h}_{k,u}^k[4] \!&\! {h}_{k,u}^k[3] \!&\! {h}_{k,u}^k[2] \\
            {h}_{k,u}^k[2] \!\!&\!\! {h}_{k,u}^k[1] \!&\! {h}_{k,u}^k[0]{+}{h}_{k,u}^k[4] \!&\!  {h}_{k,u}^k[3] \\
            {h}_{k,u}^k[3] \!\!&\!\! {h}_{k,u}^k[2] \!&\! {h}_{k,u}^k[1] \!&\! {h}_{k,u}^k[0]{+}{h}_{k,u}^k[4]\\
        \end{array}\!\right]}_{{\bf \bar H}_{k,u}^{\rm C}} {\bf f}_1s_{k,u} \nonumber \\
    &~~~+ \sum_{i\neq k}\sum_{u=1}^3 \underbrace{\left[\!\begin{array}{ccccc}
            0              \!&\! {h}_{i,u}^k[3] \!&\! {h}_{i,u}^k[2] \!&\! 0 \\
            0              \!&\! 0              \!&\! {h}_{i,u}^k[3] \!&\! {h}_{i,u}^k[2] \\
            {h}_{i,u}^k[2] \!&\! 0              \!&\! 0              \!&\! {h}_{i,u}^k[3] \\
            {h}_{i,u}^k[3] \!&\! {h}_{i,u}^k[2] \!&\! 0              \!&\! 0 \\
        \end{array}\!\right]}_{{\bf \bar H}_{i,u}^k}{\bf f}_1s_{i,u}
        + \underbrace{\left[\!\begin{array}{c}
            {z}_{k}[4] \\ {z}_{k}[1]{+}{z}_{k}[5] \\ {z}_{k}[2]{+}z_k [6] \\ {z}_{k}[7] \\
        \end{array}\!\right]}_{{\bf \bar z}_k}.
\end{align}}
\hrulefill
\end{figure*}
\fi
%\begin{align}\label{eq:ex2:rcombine}
%    &{\bf \bar y}_k = {\bf D}{\bf \bar y}_k =  \left[y_k[4], y_k[1]{+}y_k[5], y_k[2]{+}y_k[6], y_k[7]\right]^{\top} \nonumber \\
%    &= \sum_{u=1}^{3} {\bf \bar H}_{k,u}^{\rm NC}{\bf f}_{1}s_{k,u}
%        + \sum_{u=1}^{3} {\bf \bar H}_{k,u}^{\rm C}{\bf f}_{1}s_{k,u}
%        + \sum_{i\neq k}\sum_{u=1}^{3} {\bf \bar H}_{i,u}^k{\bf f}_{1}s_{i,u}
%        + {\bf \bar z}_k,
%\end{align}
%where ${\bf \bar z}_{k} =\left[z_k[4], z_k[1]{+}z_k[5], z_k[2]{+}z_k[6], z_k[7]\right]^{\top}$,
%    \begin{align}
%        {\bf \bar H}_{k,u}^{\rm C} = {\rm Circ}\Big( \Big[{h}_{k,u}^k[0]{+}{h}_{k,u}^k[4],{h}_{k,u}^k[1],{h}_{k,u}^k[2],{h}_{k,u}^k[3] \Big] \Big),
%    \end{align}
%    \begin{align}
%        {\bf \bar H}_{i,u}^k = {\rm Circ}\Big( \Big[0,0,{h}_{i,u}^k[2],{h}_{i,u}^k[3] \Big] \Big),
%    \end{align}
%    and
%    \begin{align}
%        {\bf \bar H}_{k,u}^{\rm NC} = \left[\!\begin{array}{cccc}
%            -{h}_{k,u}^k[4] \!&\! 0 \!&\! 0 \!&\! 0 \\
%            0 \!&\! {h}_{k,u}^k[0] \!&\! 0 \!&\! 0 \\
%            0 \!&\! {h}_{k,u}^k[1] \!&\! {h}_{k,u}^k[0] \!&\! 0  \\
%            0 \!&\! 0 \!&\! 0\!&\! ~~~0~~~ \\
%        \end{array}\!\right].
%    \end{align}
Therefore, we are able to eliminate all ICI signals by using a second-stage receive combining matrix ${\bf W}_2 = [{\bf f}_2,{\bf f}_3,{\bf f}_4]^H$,
    where ${\bf f}_2=0.5[1,  j , -1, j]^{\top}$, ${\bf f}_3=0.5[1,  -1 , 1, -1]^{\top}$, and ${\bf f}_4=0.5 [1,  -j , -1, j]^{\top}$.
Then the effective received signal ${\bf \tilde y}$ after the receive combining is obtained as
\ifonecol
\begin{align}\label{eq:ex2:effreceived}
    &{\bf \tilde y}_k
    = {\bf W}_2{\bf \bar y}_k
        = \sum_{u=1}^{3} {\bf W}_2{\bf \bar H}_{k,u}^{\rm NC}{\bf f}_{1}s_{k,u} + {\bf W}_2{\bf \bar z}_k
        = \underbrace{\left[\!\begin{array}{ccc}
                \frac{-1{-}j}{4} \!& \frac{-1}{4} & \frac{-1}{4} \\
                0     \!& \frac{1}{4} & \frac{-1}{4} \\
                \frac{-1{+}j}{4} \!& \frac{j}{4} & \frac{-1}{4}  \\
            \end{array}\!\right]
            \left[\!\begin{array}{ccc}
                {h}_{k,1}^k[0] \!&\! {h}_{k,2}^k[0] \!&\! {h}_{k,3}^k[0] \\
                {h}_{k,1}^k[1] \!&\! {h}_{k,2}^k[1] \!&\! {h}_{k,3}^k[1] \\
                {h}_{k,1}^k[4] \!&\! {h}_{k,2}^k[4] \!&\! {h}_{k,3}^k[4] \\
            \end{array}\!\right]}_{{\bf \tilde H}_k}
            \underbrace{\left[\!\begin{array}{c}
                s_{k,1} \\ s_{k,2} \\ s_{k,3}
            \end{array}\!\right]}_{{\bf s}_k} +  \tilde{\bf  z}_k,
\end{align}
\else
\begin{align}\label{eq:ex2:effreceived}
    &{\bf \tilde y}_k
    = {\bf W}_2{\bf \bar y}_k
        = \sum_{u=1}^{3} {\bf W}_2{\bf \bar H}_{k,u}^{\rm NC}{\bf f}_{1}s_{k,u} + {\bf W}_2{\bf \bar z}_k \nonumber \\
    &= \underbrace{-\frac{1}{4}\left[\!\begin{array}{ccc}
                1{+}j \!& 1 & 1 \\
                0     \!& -1 & 1 \\
                1{-}j \!& 1 & 1  \\
            \end{array}\!\right]
            \left[\!\begin{array}{ccc}
                {h}_{k,1}^k[0] \!&\! {h}_{k,2}^k[0] \!&\! {h}_{k,3}^k[0] \\
                {h}_{k,1}^k[1] \!&\! {h}_{k,2}^k[1] \!&\! {h}_{k,3}^k[1] \\
                {h}_{k,1}^k[4] \!&\! {h}_{k,2}^k[4] \!&\! {h}_{k,3}^k[4] \\
            \end{array}\!\right]}_{{\bf \tilde H}_k}
            \underbrace{\left[\!\begin{array}{c}
                s_{k,1} \\ s_{k,2} \\ s_{k,3}
            \end{array}\!\right]}_{{\bf s}_k} +  \tilde{\bf  z}_k,
\end{align}
\fi
where $\tilde{\bf  z}_k = {\bf W}_2{\bf \bar z}_k$.
Note that ${\bf \tilde H}_k$ in \eqref{eq:ex2:effreceived} is a full rank matrix because $h_{k,j}^k[\ell]$ is independently drawn from a continuous distribution. As a result, it is possible to reliably decode all three data symbols in ${{\bf s}_k}$ at the $k$-th BS, i.e., $d_{\Sigma}=\frac{3K}{7}$.

This result is a remarkable gain compared to that attained by the proposed method without considering the delay offset, given as $d_{\Sigma}^{\rm MAC}=K\frac{L_{\rm D}-L_{\rm I}}{\max\left\{L_{\rm D},2L_{\rm I}-1\right\}} = \frac{K}{7}$ in Corollary~1. This DoF gain stems from the additional use of ICI-free received signals during the ICI delay offset, which essentially  increases the rank of the effective channel matrix after the ICI cancellation.

\subsection{Effect of Propagation Delay When $L_{\rm D} \leq L_{\rm I}$}
In the case of $L_{\rm D} \leq L_{\rm I}$, Theorem 1 implies that without CSIT, only the trivial sum-DoF of 1 is achievable for the $K$-cell MAC with ISI.
% the derived DoF result does not hold.
Nevertheless, it is still possible to apply the proposed interference management method to obtain a spectral efficiency gain instead.  The basic idea is to consider a fraction of ICI signals as an \emph{effective ICI} signals, while treating the remaining ICI signals as additional noise. For example, we can treat the last $L_{\rm I}-L_{\rm D}$ CIR taps of ICI links as noise, so that we can apply the modified interference management method presented in the previous subsection. In the following, we demonstrate this approximation method by using an illustrative example.
%To validate this approximation method, we provide a numerical result in Example 3.

\vspace{0.1cm}
{\bf Example 3 ($L_{\rm D}=5$ and $L_{\rm I}=7$):} Consider a $K$-cell MAC with three users per cell when $L_{k,k} =L_{\rm D}=5$ for $u\in \mathcal{U}_k$ and $L_{k,i} =L_{\rm I}=7$ for $u \in \mathcal{U}_i$, $i\neq k$, and $i, k\in\mathcal{K}$. Suppose that there exist the delay-offset of three time slots for all ICI links, i.e., $L_{\rm I,d}=3$. We intentionally treat the last two CIR taps of ICI links as noise. Then the number of considered CIR taps of ICI links is given by $L_{\rm I}^\prime = 5$, and consequently, $L_{\rm I}^{\rm eff} = L_{\rm I}^\prime - L_{\rm I,d} = 2$. The considered scenario is depicted in Fig.~\ref{fig:extension}(c). In this scenario, we will show that the effect of residual ICI signals are negligible in some practical wireless environments.

Using the transmission strategy of the proposed interference management method, suppose that all users use the first column of the $5$-point IDFT matrix to convey one data symbol, i.e., ${\bf \bar x}_{k,u} = {\bf f}_1 s_{k,u} \in \mathbb{C}^{5}$, where ${\bf f}_i$ is the $i$-th column of 5-point IDFT matrix. The transmitted signal vector is generated by adding the cyclic prefix with the length of $L_{\rm I}^\prime -1 = 4$ to $\bar{\bf x}_{k,u}$. As in Example~2, we use the first-stage receive combining matrix defined as
\begin{align}
    {\bf W}_{\rm 1} = \left[\!\begin{array}{ccccccccc}
            0 & 0 & 0 & 0 & 1 & 0 & 0 & 0 & 0 \\
            1 & 0 & 0 & 0 & 0 & 1 & 0 & 0 & 0 \\
            0 & 1 & 0 & 0 & 0 & 0 & 1 & 0 & 0 \\
            0 & 0 & 1 & 0 & 0 & 0 & 0 & 1 & 0  \\
            0 & 0 & 0 & 0 & 0 & 0 & 0 & 0 & 1 \\
        \end{array}\!\right],
\end{align}
and also use ${\bf W}_{\rm 2} =\left[{\bf f}_2,{\bf f}_3,{\bf f}_4,{\bf f}_5\right]^H$ as the second-stage receive combining matrix.
Then the effective received signal ${\bf \tilde y}\in \mathbb{C}^4$ after the receive combining is obtained as
\ifonecol
\begin{align}\label{eq:ex3:effreceived}
    &{\bf \tilde y}_k
    = {\bf W}_2{\bf W}_1{\bf y}_k = \sum_{u=1}^{3} {\bf W}_2{\bf \bar H}_{k,u}^{\rm NC}{\bf f}_{1}s_{k,u} +
        \sum_{i\neq k} \sum_{u=1}^{3} {\bf W}_2{\bf \bar H}_{k,i,u}^{\rm NC}{\bf f}_{1}s_{i,u} +  {\bf W}_2{\bf W}_1{\bf z}_k,
\end{align}
\else
\begin{align}\label{eq:ex3:effreceived}
    &{\bf \tilde y}_k
    = {\bf W}_2{\bf W}_1{\bf y}_k \nonumber \\
    &= \sum_{u=1}^{3} {\bf W}_2{\bf \bar H}_{k,u}^{\rm NC}{\bf f}_{1}s_{k,u} +
        \sum_{i\neq k} \sum_{u=1}^{3} {\bf W}_2{\bf \bar H}_{k,i,u}^{\rm NC}{\bf f}_{1}s_{i,u} +  {\bf W}_2{\bf W}_1{\bf z}_k,
\end{align}
\fi
where ${\bf \bar H}_{k,i,u}^{\rm NC}$ is the non-circulant part of ${\bf \bar H}_{i,u}^{k}$.
Unlike in \eqref{eq:ex2:effreceived}, some ICI signals are remained in the effective received signal due to the ignored CIR taps of ICI links.

To clarify the characteristic of the remaining ICI signals, we specify each CIR tap by considering the distance-based large-scale fading and the exponentially-decaying power-delay profile (PDP). Let $\alpha$ be the path-loss exponent, $\beta_{k,i}$ be a constant that determines the rate of a power reduction in PDP of the wireless channel between the user in cell $i$ and the $k$-th BS. Then the $\ell$-th tap of the CIR between user $(i,u)$ and the $k$-th BS is modeled as $h_{i,u}^k[\ell] = \sqrt{P_0}d_{k,i,u}^{-\frac{\alpha}{2}}\check{h}_{i,u}^k[\ell]$, where $\sqrt{P_0}$ is the reference path loss at $1$ meter, $d_{k,i,u}$ is the distance (in meters) between user $(i,u)$ and the $k$-th BS, $\check{h}_{i,u}^k[\ell]$ is IID as $\mathcal{CN}(0,\gamma_{k,i,\ell})$, and
\begin{align}\label{eq:ex3:PDP}
    \gamma_{k,i,\ell} =
    \begin{cases}
    	\frac{{\rm e}^{-\ell\beta_{k,k}}}{\sum_{\ell=0}^{L_{\rm D}-1}{\rm e}^{-\ell\beta_{k,k}}}, &k=i~~\text{and}~~ 0 \leq \ell \leq L_{\rm D}-1, \\
    	\frac{{\rm e}^{-\ell\beta_{k,i}}}{\sum_{\ell=L_{{\rm I,d}}}^{L_{\rm I}-1}{\rm e}^{-\ell\beta_{k,i}}}, &k\neq i ~~\text{and}~~ L_{\rm I,d} \leq \ell \leq L_{\rm I}-1, \\
    	0, & \text{otherwise}.
    \end{cases}
\end{align}
Based on this model, the effective received signal in \eqref{eq:ex3:effreceived} is rewritten as
\ifonecol
\begin{align}\label{eq:ex3:effreceived2}
    {\bf \tilde y}_k
    = &\sqrt{{P}_0}\sum_{u=1}^{3} d_{k,k,u}^{-\frac{\alpha}{2}} {\bf W}_2\check{\bf H}_{k,u}^{k}{\bf f}_{1}s_{k,u} -\sqrt{{P}_0}\sum_{i\neq k} \sum_{u=1}^{3} d_{k,i,u}^{-\frac{\alpha}{2}}{\bf W}_2\check{\bf H}_{i,u}^{k}{\bf f}_{1}s_{i,u} +  {\bf W}_2{\bf W}_1{\bf z}_k,
\end{align}
\else
\begin{align}\label{eq:ex3:effreceived2}
    {\bf \tilde y}_k
    = &\sqrt{{P}_0}\sum_{u=1}^{3} d_{k,k,u}^{-\frac{\alpha}{2}} {\bf W}_2\check{\bf H}_{k,u}^{k}{\bf f}_{1}s_{k,u} \nonumber \\
    &-\sqrt{{P}_0}\sum_{i\neq k} \sum_{u=1}^{3} d_{k,i,u}^{-\frac{\alpha}{2}}{\bf W}_2\check{\bf H}_{i,u}^{k}{\bf f}_{1}s_{i,u} +  {\bf W}_2{\bf W}_1{\bf z}_k,
\end{align}
\fi
where
\ifonecol

\vspace{-3mm}{\small{\begin{align}\label{eq:ex3:H_kku}
    \check{\bf H}_{k,u}^{k} = \left[\!\begin{array}{ccccc}
            0 & 0 & 0 & 0 & 0 \\
            0 & \check{h}_{k,u}^k[0] & 0 & 0 & 0 \\
            0 & \check{h}_{k,u}^k[1] & \check{h}_{k,u}^k[0] & 0 & 0 \\
            0 & \check{h}_{k,u}^k[2] & \check{h}_{k,u}^k[1] & \check{h}_{k,u}^k[0] & 0 \\
            0 & 0 & 0 & 0 & 0 \\
        \end{array}\!\right],
~~\text{and}~~
    \check{\bf H}_{i,u}^{k} = \left[\!\begin{array}{ccccc}
            \check{h}_{i,u}^k[5] & 0 & ~~0~~ & 0 & \check{h}_{i,u}^k[6]  \\
            \check{h}_{i,u}^k[6] & 0 & 0 & 0 & 0 \\
            0 & 0 & 0 & 0 & 0 \\
            0 & 0 & 0 & 0 & 0 \\
            0 & 0 & 0 & 0 & 0 \\
        \end{array}\!\right], \nonumber
\end{align}}}\noindent
\else

\vspace{-3mm}{\small{\begin{align}
    \check{\bf H}_{k,u}^{k} = \left[\!\begin{array}{ccccc}
            0 & 0 & 0 & 0 & 0 \\
            0 & \check{h}_{k,u}^k[0] & 0 & 0 & 0 \\
            0 & \check{h}_{k,u}^k[1] & \check{h}_{k,u}^k[0] & 0 & 0 \\
            0 & \check{h}_{k,u}^k[2] & \check{h}_{k,u}^k[1] & \check{h}_{k,u}^k[0] & 0 \\
            0 & 0 & 0 & 0 & 0 \\
        \end{array}\!\right], \nonumber
\end{align}
and
\begin{align}
    \check{\bf H}_{i,u}^{k} = \left[\!\begin{array}{ccccc}
            \check{h}_{i,u}^k[5] & 0 & ~~0~~ & 0 & \check{h}_{i,u}^k[6]  \\
            \check{h}_{i,u}^k[6] & 0 & 0 & 0 & 0 \\
            0 & 0 & 0 & 0 & 0 \\
            0 & 0 & 0 & 0 & 0 \\
            0 & 0 & 0 & 0 & 0 \\
        \end{array}\!\right], \nonumber
\end{align}}}\noindent
\fi
for $i\neq k$ and $i,k\in\mathcal{K}$.
As can be seen in \eqref{eq:ex3:effreceived2}, the effect of the remaining ICI signals is negligible if $d_{k,i,u^\prime}^{\frac{\alpha}{2}} \gg d_{k,k,u}^{\frac{\alpha}{2}}$ for $u \in \mathcal{U}_k$, $u^\prime \in \mathcal{U}_i$, $i\neq k$, and $i,k \in \mathcal{K}$. This corresponds to the case that each user is far from cell edge or the path-loss exponent is high.  Furthermore, the magnitude of the remaining ICI signals exponentially decreases as $\beta_{k,i}$ increases for  $i\neq k$, and $i,k \in \mathcal{K}$.
Therefore, in such cases, each BS can effectively decode all three signals by applying well-known detection methods such as MLD and ZFD.

\begin{figure}
\centering \vspace{-0.3cm}
\includegraphics[width=3.1in]{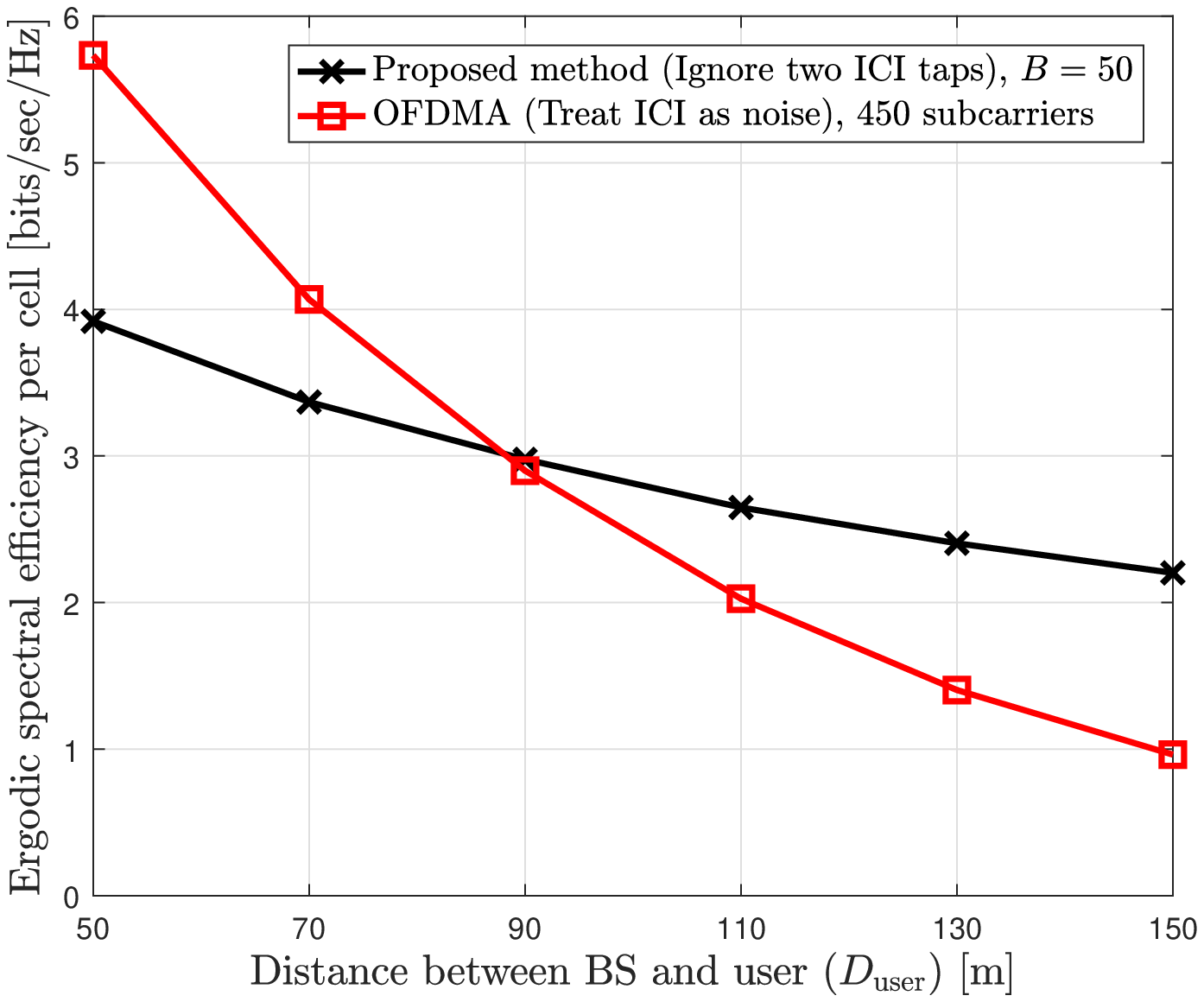} \vspace{-0.3cm}\caption{Ergodic spectral efficiencies per cell of the proposed interference management method and OFDMA for different distances ($D_{\rm user}$) between a BS and each user. We consider 7-cell deployment, as illustrated in Fig.~\ref{fig:Deploy}, in which $D_{\rm site} = 300$ m. We plot the performance of the center cell. We set $\alpha=3.5$, $P_0=-80$ dB, and $\beta_{k,i}=0.5$ for $i,k\in\mathcal{K}$. Total transmission power at the user is set to be $23$ dBm, and the noise power is set to be $-174$ dBm/Hz. Other simulation parameters are specified in Example 3.} \label{fig:Ex3}
\end{figure}

\begin{figure}
\centering \vspace{-0.3cm}
\includegraphics[width=3.1in]{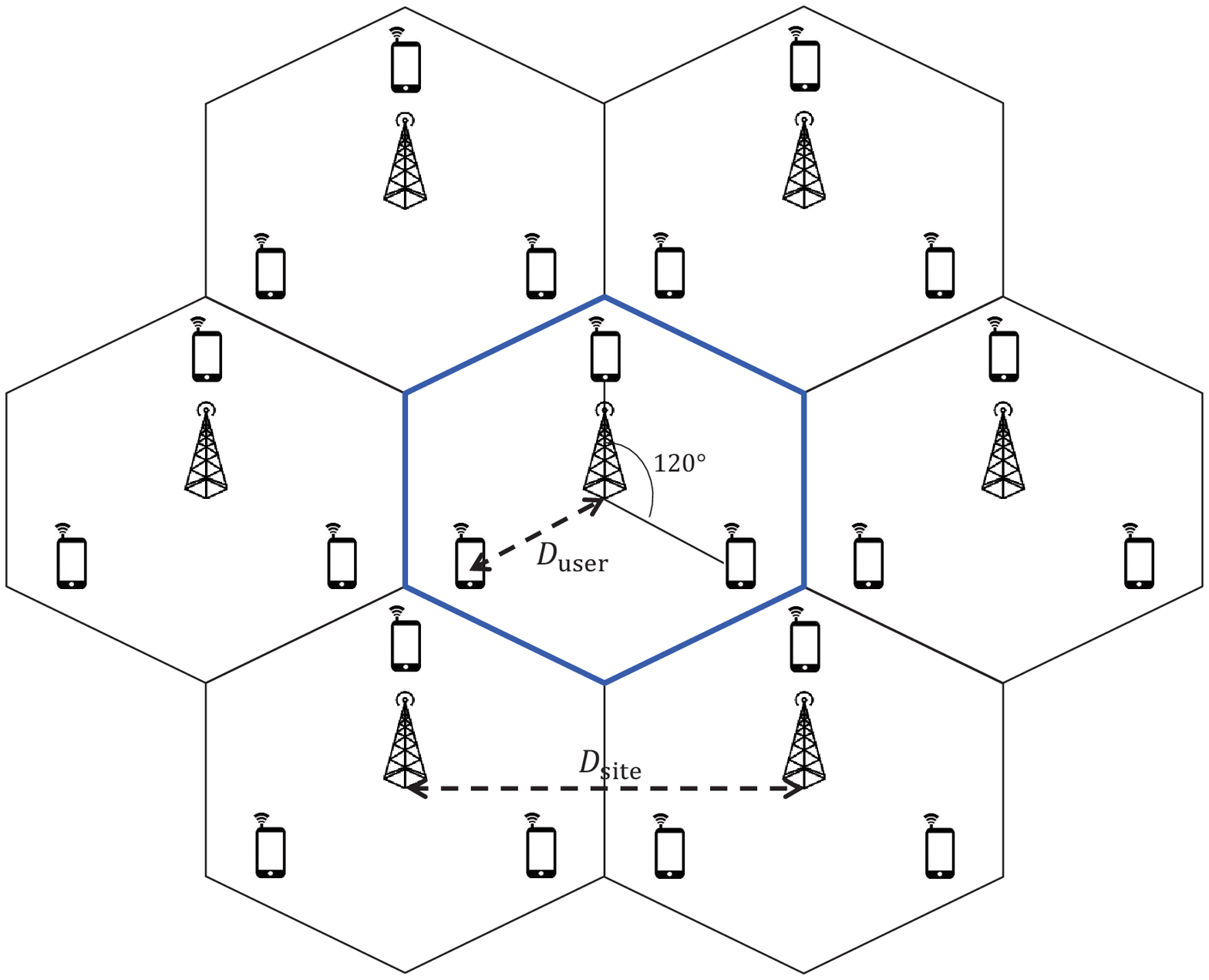} \vspace{-0.3cm}\caption{Illustration of 7-cell deployment scenario with three users per cell.} \label{fig:Deploy}
\end{figure}

In Fig. \ref{fig:Ex3}, we also validate the effectiveness of the interference management strategy presented in the above example by simulations. In this simulation, we assume that each user uses the Gaussian signaling, so the achievable spectral efficiency at the $k$-th BS sent over $T = B{\bar N} + L_{\rm I}-1$ time slots is computed as
\ifonecol
\begin{align}
    R_k= \frac{B}{T}\log_2\left|{\bf I}_5 + \left(\sum_{i\neq k}{\bf \tilde H}_{k,i}^{\rm int}({\bf \tilde H}_{k,i}^{\rm int})^H +  {\bf W}_2{\bf W}_1 {\bf W}_1^H{\bf W}_2^H\right)^{-1}{\bf \tilde H}_{k}{\bf \tilde H}_{k}^H \right|,
\end{align}
\else

\vspace{-3mm}{\small{\begin{align}
    R_k= \frac{B}{T}\log_2\left|{\bf I}_5 + \left(\sum_{i\neq k}{\bf \tilde H}_{k,i}^{\rm int}({\bf \tilde H}_{k,i}^{\rm int})^H +  {\bf W}_2{\bf W}_1 {\bf W}_1^H{\bf W}_2^H\right)^{-1}{\bf \tilde H}_{k}{\bf \tilde H}_{k}^H \right|,
\end{align}}}\noindent
\fi
where ${\bf \tilde H}_{k} = {\bf W_2}\left[{\bf \bar H}_{k,1}^{\rm NC}{\bf f}_{1},{\bf \bar H}_{k,2}^{\rm NC}{\bf f}_{1},{\bf \bar H}_{k,3}^{\rm NC}{\bf f}_{1}\right]$, and ${\bf \tilde H}_{k,i}^{\rm int} = {\bf W_2}\left[{\bf \bar H}_{k,i,1}^{\rm NC}{\bf f}_{1},{\bf \bar H}_{k,i,2}^{\rm NC}{\bf f}_{1},{\bf \bar H}_{k,i,3}^{\rm NC}{\bf f}_{1}\right]$.
%\begin{align}
%    {\bm \Sigma}_k &= \sum_{i\neq k}{\bf \tilde H}_{k,i}^{\rm int}({\bf \tilde H}_{k,i}^{\rm int})^H +  {\bf W}_2{\bf W}_1 {\bf W}_1^H{\bf W}_2^H, \nonumber \\
%    {\bf \tilde H}_{k} &= {\bf W_2}\left[{\bf \bar H}_{k,1}^{\rm NC}{\bf f}_{1},{\bf \bar H}_{k,2}^{\rm NC}{\bf f}_{1},{\bf \bar H}_{k,3}^{\rm NC}{\bf f}_{1}\right],~~\text{and} \nonumber \\
%   {\bf \tilde H}_{k,i}^{\rm int} &= {\bf W_2}\left[{\bf \bar H}_{k,i,1}^{\rm NC}{\bf f}_{1},{\bf \bar H}_{k,i,2}^{\rm NC}{\bf f}_{1},{\bf \bar H}_{k,i,3}^{\rm NC}{\bf f}_{1}\right]. \nonumber
%\end{align}
Fig. \ref{fig:Ex3} shows that when ICI is dominant, the proposed interference management method outperforms OFDMA. This implies that the proposed interference management method effectively mitigates ICI even if we ignore some CIR taps of ICI links. Whereas, when the users are close to the BS, ICI is no longer a dominant factor, so the spectral efficiency of the proposed interference management method becomes lower than that of OFDMA which treats ICI as noise. From this numerical example, we can conclude that the proposed interference management strategy can still be used as an interference-mitigation method even when $L_{\rm D} \leq L_{\rm I}$ case, and its effectiveness may heavily depend on the communicating environments.
% In all cases, the spectral efficiency of TDMA-OFDMA performs worse than both the proposed IA method and OFDMA method that treats ICI as noise.

%%%%%%%%%%%%%%%%%%%%%%%%%%%%%%%%%%%%%%%%%%%%%%%%%%%%%%%%%%%%%%%%%%

\section{Conclusion}
In this work, we showed that the interference-free sum-DoF of $K$ is asymptotically achievable in a $K$-cell SISO MAC with ISI, even in the absence of CSIT. This achievability was demonstrated by a blind interference management method that exploits the relativity in delay spreads between desired and interfering links. The result of this work is surprising because the existing work on multi-cell MAC has been shown to asymptotically achieve the sum-DoF of $K$ only when global and perfect CSIT are available \cite{Suh_Tse:08,Cadambe_Jafar:09}. We also observed that a significant DoF gain compared to the result in \cite{Lee:16} is obtained when multiple users exist in a cell by improving the utilization of signal dimensions.

One can easily show that the similar DoF gain is also achieved for a $K$-cell broadcasting channel (or interfering broadcasting channel) as this channel has a duality property with the $K$-cell MAC, so called \emph{uplink-downlink} duality. Therefore, it would be interesting to investigate the sum-DoF of the $K$-cell broadcasting channel with ISI as a future work, by applying the interference management strategy presented in this work. Another promising direction for future work is to extend the proposed interference management method for a multi-antenna setting, i.e., $K$-cell MIMO MAC with ISI. In this extension, it may be able to further improve the sum-DoF by exploiting additional signal dimensions provided by the use of multiple antennas.

\appendices
\section{Proof of Lemma~\ref{Lemma2}}
In this proof, we show that the rank of the effective channel matrix ${\tilde{\bf H}_k} \in \mathbb{C}^{(N-M_{\rm D})\times U_k^\prime M_k}$ defined in \eqref{eq:effchannel} is $U_k^\prime M_k$. Because $U_k^\prime M_k \leq L_{k,k}-L_{\rm I} \leq N-M_{\rm D}$ by the definitions, we can equivalently show that ${\tilde{\bf H}_k}$ is a full column rank matrix.

We start by introducing the \emph{rank-equivalent} representation of ${\tilde{\bf H}_k}$. Let ${\bf h}_{k,u}^{\rm eff} \in \mathbb{C}^{L_{k,k}-L_{\rm I}}$ be an effective channel-coefficient vector of user $(k,u)$, defined as ${\bf h}_{k,u}^{\rm eff} = \Big[h_{k,u}^{k}[L_{\rm I}],\cdots, h_{k,u}^{k}[L_{k,k}-1]\Big]^{\top}$. Using this vector, ${\bf W}{\bf \bar H}_{k,u}^{\rm NC}{\bf f}_{m}$ can be decomposed as follows:
\begin{align}\label{eq:basisrep}
    {\bf W}{\bf \bar H}_{k,u}^{\rm NC}{\bf f}_{m} = \underbrace{{\bf W} {\bf D}_{m,k}^{(1)} {\bf E}_k {\bf D}_{m,k}^{(2)}}_{\triangleq {\bf G}_{m,k}} {\bf h}_{k,u}^{\rm eff}.
\end{align}
In the above decomposition, ${\bf D}_{m,k}^{(1)}\in \mathbb{C}^{N}$ and ${\bf D}_{m,k}^{(2)} \in \mathbb{C}^{L_{k,k}-L_{\rm I}}$ are diagonal matrices defined as
\ifonecol
    \begin{align}
           {\bf D}_{m,k}^{(1)} &= w_m^{L_{k,k}-L_{\rm I}} {\rm diag}\Big( \Big[w_m^{L_{\rm I}-L_{k,k}},w_m^{L_{\rm I}-L_{k,k}+1},\cdots, w_m^{-2}, w_m^{-1}, \underbrace{1,1,\cdots,1}_{N-L_{k,k}+L_{\rm I}}\Big] \Big), \label{eq:Dm1}\\
%        {\bf D}_{m,k}^{(1)} &= w_m^{L_{k,k}-L_{\rm I}} {\rm diag}\Big( \Big[w_m^{L_{\rm I}-L_{k,k}},\cdots,w_m^{N-1-L_{k,k}}, \nonumber \\
%            &0, w_m^{N+1-L_{k,k}}, w_m^{N+2-L_{k,k}},\cdots,w_m^{-1}, \underbrace{1,1,\cdots,1}_{N-L_{k,k}+L_{\rm I}}\Big] \Big), \label{eq:Dm1}\\
%        {\bf D}_{m,k}^{(1)} &= {\rm diag}\Big( \big[1,w_m,\cdots,w_m^{N-L_{\rm I}-1},0, \nonumber \\
%            &\qquad\qquad w_m^{N-L_{\rm I}+1}, w_m^{N-L_{\rm I}+2},\cdots,w_m^{L_{\rm D}-L_{\rm I}+1},\nonumber \\
%            &\qquad\qquad \underbrace{w_m^{L_{\rm D}-L_{\rm I}},\cdots,w_m^{L_{\rm D}-L_{\rm I}}}_{N-L_{\rm D}+L_{\rm I}}\big] \Big) \in \mathbb{C}^{N\times N}, \label{eq:Dm1}\\
        {\bf D}_{m,k}^{(2)} &= w_m^{N-L_{\rm I}}{\rm diag}\Big( \Big[1,w_m^{-1},\cdots,w_m^{-L_{k,k}+L_{\rm I}+1} \Big] \Big), \label{eq:Dm2}
    \end{align}
\else
    \begin{align}
           {\bf D}_{m,k}^{(1)} &= w_m^{L_{k,k}-L_{\rm I}} {\rm diag}\Big( \Big[w_m^{L_{\rm I}-L_{k,k}},w_m^{L_{\rm I}-L_{k,k}+1},\cdots, \nonumber \\
            &\qquad\qquad \qquad\qquad w_m^{-2}, w_m^{-1}, \underbrace{1,1,\cdots,1}_{N-L_{k,k}+L_{\rm I}}\Big] \Big), \label{eq:Dm1}\\
%        {\bf D}_{m,k}^{(1)} &= w_m^{L_{k,k}-L_{\rm I}} {\rm diag}\Big( \Big[w_m^{L_{\rm I}-L_{k,k}},\cdots,w_m^{N-1-L_{k,k}}, \nonumber \\
%            &0, w_m^{N+1-L_{k,k}}, w_m^{N+2-L_{k,k}},\cdots,w_m^{-1}, \underbrace{1,1,\cdots,1}_{N-L_{k,k}+L_{\rm I}}\Big] \Big), \label{eq:Dm1}\\
%        {\bf D}_{m,k}^{(1)} &= {\rm diag}\Big( \big[1,w_m,\cdots,w_m^{N-L_{\rm I}-1},0, \nonumber \\
%            &\qquad\qquad w_m^{N-L_{\rm I}+1}, w_m^{N-L_{\rm I}+2},\cdots,w_m^{L_{\rm D}-L_{\rm I}+1},\nonumber \\
%            &\qquad\qquad \underbrace{w_m^{L_{\rm D}-L_{\rm I}},\cdots,w_m^{L_{\rm D}-L_{\rm I}}}_{N-L_{\rm D}+L_{\rm I}}\big] \Big) \in \mathbb{C}^{N\times N}, \label{eq:Dm1}\\
        {\bf D}_{m,k}^{(2)} &= w_m^{N-L_{\rm I}}{\rm diag}\Big( \Big[1,w_m^{-1},\cdots,w_m^{-L_{k,k}+L_{\rm I}+1} \Big] \Big), \label{eq:Dm2}
    \end{align}
\fi
respectively, where $w_m = e^{j\frac{2\pi}{N}(m-1)}$, and ${\bf E}_k \in \mathbb{C}^{N\times(L_{k,k}-L_{\rm I})}$ is a matrix with the special form:
    \begin{align}\label{eq:E_k}
        {\bf E}_k = \left[\begin{array}{cc}
            -{\bf 1}_{(N-L_{\rm I})\times (N-L_{\rm I})}^{\rm upp}   &  {\bf 0}_{(N-L_{\rm I}) \times (L_{k,k}-N)} \\
            {\bf 0}_{1 \times (N-L_{\rm I})} & {\bf 0}_{1 \times (L_{k,k}-N)}  \\
            {\bf 0}_{(L_{k,k}-N) \times (N-L_{\rm I})} &  {\bf 1}_{(L_{k,k}-N)\times(L_{k,k}-N)}^{\rm low}  \\
            {\bf 0}_{(N-L_{k,k}+L_{\rm I}-1) \times (N-L_{\rm I})} &  {\bf 1}_{(N-L_{k,k}+L_{\rm I}-1) \times (L_{k,k}-N)}  \\
        \end{array}\right].
    \end{align}
In \eqref{eq:E_k}, ${\bf 1}_{M\times M}^{\rm upp}$ and ${\bf 1}_{M\times M}^{\rm low}$ are the $M$ by $M$ upper and lower triangular matrices whose elements are all ones, respectively.
%    \begin{align}\label{eq:E_k}
%        {\bf E}_k
%        = \left[\begin{array}{cccc|cccc}
%                -1      & -1      & \cdots  & -1      & 0      & \cdots & \cdots  & ~0\\
%                0      & -1      & \cdots  & -1      & 0      & & & ~\vdots \\
%                \vdots & 0      & \ddots  & \vdots & \vdots & & & ~\vdots \\
%                \vdots &        & \ddots  & -1      & 0      & & & ~\vdots \\
%                0      & \cdots & \cdots  & 0      & 0      & \cdots & \cdots  & ~0\\
%                0      & \cdots & \cdots  & 0      & 1      & 0      & \cdots & ~0    \\
%                       &        &         &        & 1      & 1      & \ddots & ~\vdots     \\
%                \vdots &        &         & \vdots & \vdots & \vdots & \ddots & ~0     \\
%                \vdots &        &         & \vdots & 1      & 1      & \cdots & ~1    \\
%                       &        &         &        & \vdots & \vdots &        & ~\vdots \\
%               \undermat{N-L_{\rm I}}{0~      & \cdots & \cdots  & 0 &} \undermat{L_{k,k}-N}{ 1      & ~~1~~ & \cdots  & ~1}
%            \end{array}\right] %\in \mathbb{R}^{N\times (L_{k,k}-L_{\rm I})}.
%              \begin{array}{r@{}}
%     \left.\begin{array}{@{}c@{}}\null\\ \null\\ \null \\ \null\\ \null\\ \null\end{array}\right\}~~~~~{N-L_{\rm I}}~~~~ \\
%    \left.\begin{array}{@{}c@{}}\null\\ \null\\ \null\\ \null\\ \null\end{array}\right\}~~~~~{L_{k,k}{-}N}~~~ \\
%    \left.\begin{array}{@{}c@{}}\null\\ \null\\ \null\\ \null\end{array}\right\}~{N-L_{k,k}-L_{\rm I}}
%  \end{array}
%    \end{align}
By aggregating the effective channel-coefficient vectors in cell $k$, we also define the following random matrix:
\begin{align}\label{eq:eff_rand_matrix}
    {\bf H}_{k}^{\rm eff} = \left[ {\bf h}_{k,1}^{\rm eff},{\bf h}_{k,2}^{\rm eff},\cdots, {\bf h}_{k,U_k^\prime}^{\rm eff}\right]  \in \mathbb{C}^{(L_{k,k}-L_{\rm I})\times U_k^\prime}.
\end{align}
From \eqref{eq:basisrep} and \eqref{eq:eff_rand_matrix}, we obtain the rank-equivalent representation of ${\tilde{\bf H}_k}$ as follows:
\ifonecol
\begin{align}\label{eq:span_eff}
    {\rm rank} ({\tilde{\bf H}_k})
    &= {\rm rank}\bigg(\Big[
            {\bf W}{\bf \bar H}_{k,1}^{\rm NC}{\bf F}_{k},{\bf W}{\bf \bar H}_{k,2}^{\rm NC}{\bf F}_{k},\cdots,{\bf W}{\bf \bar H}_{k,U_k^\prime}^{\rm NC}{\bf F}_k
        \Big]\bigg) \nonumber \\
    &= {\rm rank}\bigg(
        \Big[{\bf G}_{1,k}{\bf h}_{k,1}^{\rm eff},\cdots,
        {\bf G}_{M_k,k}{\bf h}_{k,1}^{\rm eff},\cdots, {\bf G}_{1,k}{\bf h}_{k,U_k^\prime}^{\rm eff},\cdots,
        {\bf G}_{M_k,k}{\bf h}_{k,U_k^\prime}^{\rm eff} \Big]\bigg) \nonumber \\
    &= {\rm rank}\bigg(
        \Big[{\bf G}_{1,k}{\bf H}_{k}^{\rm eff},{\bf G}_{2,k}{\bf H}_{k}^{\rm eff},\cdots,{\bf G}_{M_k,k}{\bf H}_{k}^{\rm eff}\Big]\bigg),
\end{align}
\else
\begin{align}\label{eq:span_eff}
    {\rm rank} ({\tilde{\bf H}_k})
    &= {\rm rank}\bigg(\Big[
            {\bf W}{\bf \bar H}_{k,1}^{\rm NC}{\bf F}_{k},{\bf W}{\bf \bar H}_{k,2}^{\rm NC}{\bf F}_{k},\cdots,{\bf W}{\bf \bar H}_{k,U_k^\prime}^{\rm NC}{\bf F}_k
        \Big]\bigg) \nonumber \\
    &= {\rm rank}\bigg(
        \Big[{\bf G}_{1,k}{\bf h}_{k,1}^{\rm eff},\cdots,
        {\bf G}_{M_k,k}{\bf h}_{k,1}^{\rm eff},\cdots, \nonumber \\
    &\qquad \qquad~~{\bf G}_{1,k}{\bf h}_{k,U_k^\prime}^{\rm eff},\cdots,
        {\bf G}_{M_k,k}{\bf h}_{k,U_k^\prime}^{\rm eff} \Big]\bigg) \nonumber \\
    &= {\rm rank}\bigg(
        \Big[{\bf G}_{1,k}{\bf H}_{k}^{\rm eff},{\bf G}_{2,k}{\bf H}_{k}^{\rm eff},\cdots,{\bf G}_{M_k,k}{\bf H}_{k}^{\rm eff}\Big]\bigg),
\end{align}
\fi
where the last equality is obtained by an appropriate change of the columns. With this representation, we will show that
\ifonecol
\begin{align}\label{eq:span_eff2}
    {\rm rank} ({\tilde{\bf H}_k})
    = {\rm rank}\bigg(
        \Big[{\bf G}_{1,k}{\bf H}_{k}^{\rm eff},{\bf G}_{2,k}{\bf H}_{k}^{\rm eff},\cdots,{\bf G}_{M_k,k}{\bf H}_{k}^{\rm eff}\Big]\bigg) = U_k^\prime M_k,
\end{align}
\else
\begin{align}\label{eq:span_eff2}
    &{\rm rank} ({\tilde{\bf H}_k}) \nonumber \\
    &= {\rm rank}\bigg(
        \Big[{\bf G}_{1,k}{\bf H}_{k}^{\rm eff},{\bf G}_{2,k}{\bf H}_{k}^{\rm eff},\cdots,{\bf G}_{M_k,k}{\bf H}_{k}^{\rm eff}\Big]\bigg) = U_k^\prime M_k,
\end{align}
\fi
with probability one.

To proof the last equality in \eqref{eq:span_eff2}, first, we need to show that the columns of ${\bf G}_{m_1,k}{\bf H}_{k}^{\rm eff}$ are linearly independent from the columns of ${\bf G}_{m_2,k}{\bf H}_{k}^{\rm eff}$ when $m_1 \neq m_2$ and $m_1, m_2 \in\{1,2,\ldots M_k\}$. For this, it is sufficient to show that
\ifonecol
\begin{align}\label{eq:indepGh}
    {\rm Pr}\bigg({\rm span}\Big({\bf G}_{m,k}{\bf h}_{k,u}^{\rm eff} \Big) \subset
        {\rm span}\Big(\Big[{\bf G}_{1,k}{\bf h}_{k,u}^{\rm eff},\cdots,{\bf G}_{m-1,k}{\bf h}_{k,u}^{\rm eff},{\bf G}_{m+1,k}{\bf h}_{k,u}^{\rm eff},\cdots,{\bf G}_{M_k,k}{\bf h}_{k,u}^{\rm eff} \Big]\Big)\bigg) = 0,
\end{align}
\else
\begin{align}\label{eq:indepGh}
    {\rm Pr}\bigg({\rm span}\Big({\bf G}_{m,k}{\bf h}_{k,u}^{\rm eff} \Big) \subset
    &{\rm span}\Big(\Big[{\bf G}_{1,k}{\bf h}_{k,u}^{\rm eff},\cdots,{\bf G}_{m-1,k}{\bf h}_{k,u}^{\rm eff},  \nonumber \\
    &~~~~{\bf G}_{m+1,k}{\bf h}_{k,u}^{\rm eff},\cdots,{\bf G}_{M_k,k}{\bf h}_{k,u}^{\rm eff} \Big]\Big)\bigg) = 0,
\end{align}
\fi
because it is obvious that the effective channel-coefficient vectors for different users are linearly independent with probability one as their elements are independently drawn from a continuous distribution. Furthermore, the sufficient condition of \eqref{eq:indepGh} is simply given by
\ifonecol
\begin{align}\label{eq:rank_WHF}
    \text{rank}\left(\left[{\bf G}_{1,k}{\bf h}_{k,u}^{\rm eff}, {\bf G}_{2,k}{\bf h}_{k,u}^{\rm eff}, \cdots, {\bf G}_{M_k,k}{\bf h}_{k,u}^{\rm eff}\right]\right)= \text{rank}({\bf W}{\bf \bar H}_{k,u}^{\rm NC}{\bf F}_{k}) = M_k,
\end{align}
\else
\begin{align}\label{eq:rank_WHF}
    &\text{rank}\left(\left[{\bf G}_{1,k}{\bf h}_{k,u}^{\rm eff}, {\bf G}_{2,k}{\bf h}_{k,u}^{\rm eff}, \cdots, {\bf G}_{M_k,k}{\bf h}_{k,u}^{\rm eff}\right]\right)  \nonumber \\
    &= \text{rank}({\bf W}{\bf \bar H}_{k,u}^{\rm NC}{\bf F}_{k}) = M_k,
\end{align}
\fi
where the first equality is obtained from \eqref{eq:basisrep}.
Therefore, to proof \eqref{eq:indepGh}, we instead show the equation in  \eqref{eq:rank_WHF} by using the following lemma:
\vspace{0.1cm}
\begin{lem}\label{Lemma3}
    For a matrix ${\bf A}{\bf B}{\bf C}$ defined by the product of three matrices ${\bf A}$, ${\bf B}$, and ${\bf C}$, the following inequality holds:
    \begin{align}\label{eq:rank_ineq}
        \text{rank}({\bf A}{\bf B}) + \text{rank}({\bf B}{\bf C}) \leq \text{rank}({\bf B}) + \text{rank}({\bf A}{\bf B}{\bf C}).
    \end{align}
\end{lem}
\begin{IEEEproof}
    See \cite{Horn:13}.
\end{IEEEproof}
Lemma 3 implies that if we determine the ranks of three matrices ${\bf W}{\bf \bar H}_{k,u}^{\rm NC}$, ${\bf \bar H}_{k,u}^{\rm NC}{\bf F}_{k}$, and ${\bf \bar H}_{k,u}^{\rm NC}$, we can directly obtain an inequality condition for the rank of ${\bf W}{\bf \bar H}_{k,u}^{\rm NC}{\bf F}_{k}$ from \eqref{eq:rank_ineq}. Fortunately, the rank of  ${\bf \bar H}_{k,u}^{\rm NC}$ is easily determined as
\begin{align}
    \text{rank}({\bf \bar H}_{k,u}^{\rm NC}) &= L_{k,k}-L_{\rm I}, \label{eq:rank_Hnc}
\end{align}
by the definition of  ${\bf \bar H}_{k,u}^{\rm NC}$ given in \eqref{eq:noncirculantH}.
Furthermore, the ranks of ${\bf W}{\bf \bar H}_{k,u}^{\rm NC}$ and ${\bf \bar H}_{k,u}^{\rm NC}{\bf F}_{k}$ can be determined by using the property of the submatrix of the DFT matrix: Any $N_2$ columns of an $N_1$ by $N$ submatrix of the $N$-point DFT matrix, constructed by removing $N-N_1$  consequtive rows from the original DFT matrix, are linearly independent when $N - N_1 \geq N_2$; this property can easily be shown by extending the result in \cite{Rath2004} (see Appendix C in \cite{Rath2004}).
Because ${\bf W}= \left[{\bf f}_{M_{\rm D}+1},\cdots,{\bf f}_{N} \right]^H$ and  ${\bf F}_k^H =  \left[{\bf f}_{1},{\bf f}_{2},\cdots,{\bf f}_{M_k} \right]^H $ are the submatrices of the $N$-point DFT matrix, the above property implies that any $L_{k,k}-L_{\rm I}$ columns of ${\bf W}$ are linearly independent, and also that any $M_k$ columns of ${\bf F}_k^H$ are linearly independent. Using these facts along with the definition of ${\bf \bar H}_{k,u}^{\rm NC}$, we have
\begin{align}
    \text{rank}({\bf W}{\bf \bar H}_{k,u}^{\rm NC}) &= L_{k,k}-L_{\rm I}  \label{eq:rank_WH}\\
    \text{rank}({\bf \bar H}_{k,u}^{\rm NC}{\bf F}_{k}) &= \text{rank}({\bf F}_{k}^H({\bf \bar H}_{k,u}^{\rm NC})^H) = M_k.  \label{eq:rank_HF}
\end{align}
Plugging \eqref{eq:rank_Hnc}, \eqref{eq:rank_WH}, and \eqref{eq:rank_HF} to \eqref{eq:rank_ineq} yields $ M_k \leq \text{rank}({\bf W}{\bf \bar H}_{k,u}^{\rm NC}{\bf F}_{k})$. Because ${\bf W}{\bf \bar H}_{k,u}^{\rm NC}{\bf F}_{k}$ is a tall matrix with $M_k$ columns, this rank inequality directly results in \eqref{eq:rank_WHF}.

Now, to proof the last equality in \eqref{eq:span_eff2}, we only need to show that the rank of ${\bf G}_{m,k}{\bf H}_{k}^{\rm eff}$ is $U_k$. Because all elements of ${\bf H}_{k}^{\rm eff}$ are independent random variables drawn from a continuous distribution, it is sufficient to show that
\begin{align} \label{eq:rankG}
    {\rm rank}\left({\bf G}_{m,k}\right)=L_{k,k}-L_{\rm I},~~\text{for}~m\in \{1,\ldots,M_k\}.
\end{align}
Any diagonal matrix does not change the linear independence of the columns, so we have
\begin{align} \label{eq:spanWD}
    &{\rm span} \left({\bf G}_{m,k}\right) = {\rm span} \left({\bf W}{\bf D}_{m,k}^{(1)}{\bf E}_k{\bf D}_{m,k}^{(2)}\right)
        = {\rm span} \left({\bf W}{\bf D}_{m,k}^{(1)}{\bf E}_k\right).
\end{align}
\ifonecol
Let $\tilde{\bf w}_{m,k}$ be the $k$-th column of ${\bf W}{\bf D}_{m,k}^{(1)}$. Then by the definition of ${\bf E}_k$ given in \eqref{eq:E_k}, we have
\begin{align} \label{eq:spanWD2}
    &{\rm span} \left({\bf W}{\bf D}_{m,k}^{(1)}{\bf E}_k\right)  = \!{\rm span} \Big(\Big[\! \underbrace{\tilde{\bf w}_{m,1},{\cdots}, \tilde{\bf w}_{m,N-L_{\rm I}},
        \tilde{\bf w}_{m,N-L_{\rm I}+2},{\cdots}, \tilde{\bf w}_{m,L_{k,k}-L_{\rm I}}
        , \tilde{\bf w}_{m,k}^{\rm sum}}_{\triangleq {\bf G}_{m,k}^{\rm eq}} \!\Big] \Big),
\end{align}
where $\tilde{\bf w}_{m,k}^{\rm sum} = \sum_{k=L_{k,k}-L_{\rm I}+1}^{N}\!\tilde{\bf w}_{m,k}$.
\else
Let $\tilde{\bf w}_{m,k}$ be the $k$-th column of ${\bf W}{\bf D}_{m,k}^{(1)}$. Then by the definition of ${\bf E}_k$ given in \eqref{eq:E_k}, we have

\vspace{-3mm}{\small{\begin{align} \label{eq:spanWD2}
    &{\rm span} \left({\bf W}{\bf D}_{m,k}^{(1)}{\bf E}_k\right)   \nonumber \\
    &= \!{\rm span} \Big(\Big[\! \underbrace{\tilde{\bf w}_{m,1},{\cdots}, \tilde{\bf w}_{m,N-L_{\rm I}},
        \tilde{\bf w}_{m,N-L_{\rm I}+2},{\cdots}, \tilde{\bf w}_{m,L_{k,k}-L_{\rm I}}
        , \tilde{\bf w}_{m,k}^{\rm sum}}_{\triangleq {\bf G}_{m,k}^{\rm eq}} \!\Big] \Big).
\end{align}}}\noindent
where $\tilde{\bf w}_{m,k}^{\rm sum} = \sum_{k=L_{k,k}-L_{\rm I}+1}^{N}\!\tilde{\bf w}_{m,k}$.
\fi
We have already shown that any $L_{k,k}-L_{\rm I}$ columns of ${\bf W}$ are linearly independent, so the first $L_{k,k}-L_{\rm I}-1$ columns of ${\bf G}_{m,k}^{\rm eq}$ are linearly independent.
In addition, the last column of ${\bf G}_{m,k}^{\rm eq}$ is also linearly independent from other columns because $\tilde{\bf w}_{m,k}^{\rm sum}$ contains at least one vector (e.g., saying $\tilde{\bf w}_{m,L_{k,k}-L_{\rm I}+1}$) that is independent from them. Since all columns of ${\bf G}_{m,k}^{\rm eq}$ are linearly independent, \eqref{eq:rankG} holds.

Aggregating the above results implies that all columns of ${\bf G}_{1,k}{\bf H}_{k}^{\rm eff}, {\bf G}_{2,k}{\bf H}_{k}^{\rm eff}, \ldots, {\bf G}_{M_k,k}{\bf H}_{k}^{\rm eff}$ are linearly independent with probability one, so we arrive at the result in \eqref{eq:span_eff2}; this completes the proof.


\begin{thebibliography}{1}
\bibitem{Jeon:17}
Y.-S. Jeon, N. Lee, and R. Tandon, ``On the degrees of freedom of wide-band multi-cell multiple access channels with no CSIT,'' {\em to appear in Proc. IEEE Int. Symp. Inf. Theory (ISIT)}, June 2017.


%%%%% OFDMA related
\bibitem{OFDM:66}
R. W. Chang, ``Synthesis of band-limited orthogonal signals for multichannel data transmission,''
    {\em Bell Syst. Tech. J.}, vol. 45, pp. 1775--1796. Dec. 1966.

\bibitem{Weinstein:71}
S. B. Weinstein and P. M. Ebert, ``Data transmission by frequency division multiplexing using the discrete Fourier transform,''
    {\em IEEE Trans. Commun. Tech.,} vol. 19, no. 5, pp. 628--634, Oct. 1971.

\bibitem{Bingham:90}
J. A. C. Bingham, ``Multicarrier modulation for data transmission: An idea whose time has come,''
    {\em IEEE Commun. Mag.,} vol. 28, no. 5, pp. 17--25, Mar. 1990.

%\bibitem{Hirt_Massey:88}
%W. Hirt and J. L. Massey, ``Capacity of the discrete-time Gaussian channel with intersymbol interference,''
%    {\em IEEE Trans. Inf. Theory,} vol. 34, no. 3, pp. 380--388, May 1988.

\bibitem{Verdu:93}
R. S. Cheng and S. Verd\'{u}, ``Gaussian multiaccess channels with ISI: Capacity region and multiuser water-filling,''
    {\em IEEE Trans. Inf. Theory,} vol. 39, no. 3, pp. 773--786, May 1993.
%%
%\bibitem{Tse:97}
%D. Tse, ``Optimal power allocation over parallel Gaussian broadcast channels,'' {\em in Proc. IEEE ISIT}, pp. 27, Jul 1997.
%

\bibitem{Wong:99}
C. Y. Wong, R. S. Cheng, K. B. Letaief, and R. D. Murch, ``Multicarrier OFDM with adaptive subcarrier, bit, and power allocation,''
    {\em IEEE J. Sel. Areas Commun.,} vol. 17, no. 10, pp. 1747--1758, Oct. 1999.

\bibitem{Rhee_Cioffi:00}
W. Rhee and J. M. Cioffi, ``Increasing in capacity of multiuser OFDM system using dynamic subchannel allocation,''
    in {\em Proc. IEEE Int. Veh. Tech. Conf. (VTC),} vol. 2, pp. 1085--1089, May 2000.


%%%%%%%%% CoMP and Coordinated BF
%%%%% COMP
\bibitem{Gesbert:10}
D. Gesbert, S. Hanly, H. Huang, S. S. Shitz, O. Simeone, and W. Yu, ``Multi-cell MIMO cooperative networks: A new look at interference,''
    {\em IEEE J. Sel. Areas Commun.,} vol. 28, no. 9, pp. 1380--1408, Dec. 2010.

\bibitem{Lee_Sayana:11}
D. Lee, H. Seo, B. Clerckx, E. Hardouin, D. Mazzarese, S. Nagata, and K. Sayana,
``Coordinated multipoint transmission and reception in LTE-Advanced: Deployment scenarios and operational challenges,''
    {\em IEEE Commun. Mag.,} vol. 50, no. 2, pp. 148--155, Feb. 2011.

\bibitem{Clerckx_Kim:13}
B. Clerckx, H. Lee, Y.-J. Hong, and G. Kim, ``A practical cooperative multi-cell MIMO-OFDMA network based on rank coordination,''
    {\em IEEE Trans. Wireless Commun.,} vol. 12, no. 4, pp. 1481--1491, Apr. 2013.

%\bibitem{Yu_Shin:13}
%W. Yu, T. Kwon, and C. Shin, ``Multicell coordination via joint scheduling, beamforming, and power spectrum adaptation,''
%    {\em IEEE Trans. Wireless Commun.,} vol. 12, no. 7, pp. 3300--3313, July 2013.
%


%%%%% Coordinated BF
\bibitem{Dahrouj_Yu:10}
H. Dahrouj and W. Yu, ``Coordinated beamforming for the multicell multi-antenna wireless system,''
    {\em IEEE Trans. Wireless Commun.,} vol. 9, no. 5, pp. 1748--1759, May 2011.

\bibitem{Lee_Heath:15}
N. Lee, D. Morales-Jimenez, A. Lozano, and R. W. Heath, Jr., ``Spectral efficiency of dynamic coordinated beamforming: A stochastic geometry approach,''
    {\em IEEE Trans. Wireless Commun.,} vol. 14, no. 1, pp. 230--241, Jan. 2015.


%% Interference management techiniques
%%%%% Interence alignment
\bibitem{Cadambe_Jafar:08}
V. R. Cadambe and S. A. Jafar, ``Interference alignment and the degrees of freedom of the $K$ user interference channel,''
    {\em IEEE Trans. Inf. Theory,} vol. 54, no. 8, pp. 3425--3441, Aug. 2008.
%
%\bibitem{Jafar_Shamai:08}
%S. A. Jafar and S. Shamai, ``Degrees of freedom region of MIMO $X$ channel,''
%    {\em IEEE Trans. Inf. Theory,} vol. 54, no. 1, pp. 151--170, Jan. 2008.

\bibitem{Cadambe_Jafar:09}
V. R. Cadambe and S. A. Jafar, ``Interference alignment and the degrees of freedom of wireless $X$ networks,''
    {\em IEEE Trans. Inf. Theory,} vol. 55, no. 9, pp. 3893--3908, Sep. 2009.

\bibitem{Suh_Tse:08}
C. Suh and D. Tse, ``Interference alignment for cellular networks,''
    in {\em Proc. 46th Annual Allerton Conf. Commun. Control Comput. (Allerton),} Sep. 2008.

\bibitem{Chaaban:11}
A. Chaaban, A. Sezgin, B. Bandemer, and A. Paulraj, ``On Gaussian multiple access channels with interference: Achievable rates and upper bounds,''
    in {\em  Proc. 4th Int. Workshop Multiple Access Commun. (MACOM),} Sep. 2011.

%\bibitem{Grokop_Tse:11}
%L. H. Grokop, D. N. C. Tse, and R. D. Yates, ``Interference alignment for line-of-sight channels,''
%    {\em IEEE Trans. Inf. Theory,} vol. 57, no. 9, pp. 5820--5839, Sep. 2011.


%\bibitem{Cadambe_Jafar:09}
%V. R. Cadambe and S. A. Jafar, ``Parallel Gaussian interference channels are not always separable,'' {\em IEEE Trans, on Info.Theory}, vol. 55, pp. 3983--3990, Sep. 2009.


%%%%% IA for Interfering MAC
\bibitem{Kim_Love:11}
T. Kim, D. J. Love, B. Clerckx, and D. Hwang, ``Spatial degrees of freedom of the multi-cell MIMO multiple access channel,''
    in {\em Proc. IEEE Global Commun. Conf. (GLOBECOM),} Dec. 2011.

\bibitem{Lee_Shin:12}
N. Lee, W. Shin, R. W. Heath, Jr., and B. Clerckx, ``Interference alignment with limited feedback for two-cell interfering MIMO-MAC,''
    in {\em Proc. IEEE Int. Symp. Wireless Commun. Syst. (ISWCS),} pp. 566--570, Aug. 2012.

\bibitem{Yang_Paulraj:13}
H. J. Yang, W.-Y. Shin, B. C. Jung, and A. Paulraj, ``Opportunistic interference alignment for MIMO interfering multiple-access channels,''
    {\em IEEE Trans. Wireless Commun.,} vol. 12, no. 5, pp. 2180--2192, May 2013.



%%%%% limited CSIT, blind IA, IF-OFDM
% limited CSIT from feedback
\bibitem{Zhu_Guo:12}
Y. Zhu and D. Guo, ``The degrees of freedom of isotropic MIMO interference channels without state information at the transmitters,''
    {\em IEEE Trans. Inf. Theory,}  vol. 58, no. 1, pp. 341--352, Jan. 2012.

%\bibitem{Huang:12}
%C. Huang, S. A. Jafar, S. Shamai, and S. Vishwanath, ``On degrees of freedom region of MIMO networks without channel state information at transmitters,''
%    {\em IEEE Trans. on Inf. Theory,}  vol. 58, no. 2, pp. 849--857, Feb. 2012.
%

\bibitem{Maddah-Ali:12}
M. A. Maddah-Ali and D. Tse, ``Completely stale transmitter channel state information is still very useful,''
    {\em IEEE Trans. Inf. Theory,}  vol. 58, no. 7, pp. 4418--4431, July 2012.

\bibitem{Gou_Jafar:12} T. Gou and S. A. Jafar, ``Optimal use of current and outdated channel state information: Degrees of freedom of the MISO BC with mixed CSIT,''
    {\em IEEE Commun. Lett.,} vol. 16, no. 7, pp. 1084--1087, July 2012.

\bibitem{Lee_Heath:12}
N. Lee and R. W. Heath, Jr., ``Not too delayed CSIT achieves the optimal degrees of freedom,''
    in {\em Proc. 50th Annual Allerton Conf. Commun. Control Comput. (Allerton),} Oct. 2012.

\bibitem{Lee_Heath:14}
N. Lee and R. W. Heath, Jr., ``Space-time interference alignment and degrees of freedom regions for the MISO broadcast channel with periodic CSI feedback,''
    {\em  IEEE Trans. Inf. Theory,} vol. 60, no. 1, pp. 515--528, Jan. 2014.

\bibitem{Tandon:13}
R. Tandon, S. A. Jafar, S. Shamai, and H. V. Poor, ``On the synergistic benefits of alternating CSIT for the MISO-BC,''
    {\em IEEE Trans. Inf. Theory,} vol. 59, no. 7, pp. 4106--4128, July 2013.

\bibitem{Jafar_Index_coding} S. A. Jafar, ``Topological interference management through index coding,''
    {\em IEEE Trans. Inf. Theory,}  vol. 60, no. 1, pp. 529--568, Jan. 2014.



\bibitem{Vaze:12} C. S. Vaze and M. K. Varanasi, ``The degrees of freedom regions of MIMO broadcast, interference, and cognitive radio channels with no CSIT,''
    {\em IEEE Trans. Inf. Theory,} vol. 58, no. 8, pp. 5354--5374, Aug. 2012.

\bibitem{Vaze:12-2} C. S. Vaze and M. K. Varanasi, ``A new outer bound via interference localization and the degrees of freedom regions of MIMO interference networks with no CSIT,''
    {\em IEEE Trans. Inf. Theory,} vol. 58, no. 11, pp. 6853--6869, Nov. 2012.


\bibitem{Jafar:10} S. A. Jafar, ``Interference alignment --- A new look at signal dimensions in a communication network,''
    {\em Foundations and Trends in Commun. and Inf. Theory,} vol. 7, no. 1, pp. 1--134,  2011.



% Blind IA
\bibitem{Wang_Jafar:11}
C. Wang, T. Gou, and S. A. Jafar, ``Aiming perfectly in the dark- blind interference alignment through staggered antenna switching,''
    {\em IEEE Trans. Signal Process.,} vol. 59, no. 6, pp. 2734--2744, June 2011.

\bibitem{Jafar:12}
S. A. Jafar, ``Blind interference alignment,''
    {\em IEEE J. Sel. Topics Signal Process.,} vol. 6, no. 3, pp. 216--227, June 2012.

\bibitem{Jafar_Armada:15}
M. Morales-C\'{e}spedes, J. Plata-Chaves, D. Toumpakaris, S. A. Jafar, and A. G. Armada, ``Blind interference alignment for cellular networks,"
    {\em IEEE Trans. Signal Process.,} vol. 63, no. 1, pp. 41--56, Jan. 2015.

% IF-OFDM
\bibitem{Lee:16}
N. Lee, ``Interference-free OFDM: Rethinking OFDM for interference networks with inter-symbol interference,"
    arXiv:1609.02517 [cs.IT], Sep. 2016. [Online]. Available: http://arxiv.org/abs/1609.02517



%\bibitem{Jafar:12}
%S. A. Jafar, ``Blind interference alignment,''
%    {\em IEEE J. Sel. Topics Signal Process.,} vol. 6, no. 3, pp. 216--227, June 2012.
%
%
%%%%%%% Main

\bibitem{Golub:96}
G. H. Golub and C. F. V. Loan, ``Matrix computations,''
    {\em 3rd ed. The Johns Hopkins University Press}, 1996.

% Book - rank inequality
\bibitem{Horn:13}
R. A. Horn and C. R. Johnson, ``Matrix analysis,''
    {\em 2nd ed. Cambridge University Press}, 2013.



\bibitem{Rath2004}
G. Rath and C. Guillemot, ``Frame-theoretic analysis of DFT codes with erasures,''
    {\em IEEE Trans. Signal Process.,} vol. 52, no. 2, pp. 447--460, Feb. 2004.




%%%%%% Conclusion
%\bibitem{Lozano_andrew_heath} A. Lozano, J. G. Andrews, and R. W. Heath, Jr.,``Fundamental limits of cooperation,''
%    {\em  IEEE Trans. Inf. Theory,} vol. 59, no. 9, pp. 5213--5226, Sep. 2013.
%
%
%\bibitem{Lee_David_Lozano_heath} N. Lee, D. Morales-Jimenez, A. Lozano, and R. W. Heath, Jr., ``Spectral efficiency of dynamic coordinated beamforming: A stochastic geometry approach,''
%    {\em IEEE Trans. Wireless Commun.,} vol. 14, no. 1, pp. 230--241, Jan. 2015.
%
%
%
% \bibitem{Jindal_Goldsmith:04}
%N. Jindal, S. Vishwanath, and A. Goldsmith, ``On the duality of Gaussian
%multiple-access and broadcast channels,'' {\em IEEE Transactions on Information Theory}, vol. 50, no.5, pp. 768-783, May 2004.
%
%\bibitem{Yu:07}
%W. Yu, ``Multiuser water-filling in the presence of crosstalk,'' {\em  in Proc. Inform.
%Theory and Appl. Workshop (ITA)}, San Diego, U.S.A., Jan. 2007.
%
%  \bibitem{Cioffi:07}
%S. T. Chung and J. Cioffi, ``The capacity region of frequency-selective gaussian interference channels under strong interference,'' {\em IEEE Transactions on Information Theory}, vol. 55, no. 9, pp. 1812-1821, Sept. 2007.


%\bibitem{Jafar:05}
% S. A. Jafar and A. J. Goldsmith,
%\newblock ``Isotropic fading vector broadcast channels: The scalar upper bound and loss in degrees of freedom,''
%\newblock {\em IEEE Transactions on Information Theory,}  vol. 51, no. 3, pp. 848-857, 2005.


%\bibitem{Vaze:12-2}
%C. S. Vaze and M. K. Varanasi,
%\newblock ``A new outer bound via interference localization and the degrees of freedom regions of MIMO interference networks with no CSIT,''
%\newblock {\em IEEE Transactions on Information Theory,}  vol. 58, no. 11, pp. 6853-6869, Nov. 2012.


%\bibitem{Yang}
%H. Yang, W. Shin, and J. Lee,
%\newblock ``Dynamic supersymbol design of blind interference alignment for $K$-user MISO broadcast channels,''
%\newblock {\em in Proc. IEEE International Conference on Communications (ICC)}, pp. 2301-2306, 2015.
%
%
%\bibitem{Yang2}
%H. Yang, W. Shin, and J. Lee,
%\newblock ``Degrees of freedom for $K$-user SISO interference channels with blind interference alignment,''
%\newblock {\em in Proc. IEEE Asilomar Conf. on Signals, Systems, and Computers}, Nov. 2015.


%\bibitem{Yang3}
%H. Yang, W. Shin, and J. Lee,
%\newblock ``Hierarchical blind interference alignment over interference networks with finite coherence time,''
%\newblock {\em IEEE Transactions on Signal Processing}, vol. 64, no. 5, pp. 1289-1304, Mar. 2016.

%\bibitem{Wang2}
%C. Wang,
%\newblock ``Degrees of freedom characterization: the 3-user SISO interference channel with blind interference alignment,''
%\newblock {\em IEEE Commun. Lett.}, vol. 18, no. 5, pp. 757-760, May 2014.
%
%
%\bibitem{Johnny}
%M. Johnny and M. R. Aref,
%\newblock ``Sum degrees of freedom of the $K$-user interference channel with blind CSI,''
%\newblock {\em Submitted to IEEE Transactions on Wireless Communications}, [available: arXiv:1602.03328].
%
%
%
%\bibitem{Letaief}
%Y. Lu, W. Zhang, and K. Letaief,
%\newblock ``Blind interference alignment with diversity in $K$-user interference channels,''
%\newblock {\em IEEE Transactions on Communications}, vol. 62, no. 8, pp. 2850-2859, 2014.
%
%

%\bibitem{Gesbert2013}
%P. de Kerret and D. Gesbert,
%\newblock ``Interference alignment with incomplete CSIT sharing,'' \newblock {\em IEEE Trans. Wirel. Commun.,} vol. 13, no. 5, pp. 2563 - 2573, May 2014.
%
%\bibitem{Chen_Elia:13}
%J. Chen and P. Elia, ``Toward the performance versus feedback tradeoff for the two-user MISO broadcast channel,'' {\em IEEE Transactions on Information Theory}, vol. 59,  no. 12, pp. 8336-8356, Dec. 2013.
%
%
%
%\bibitem{Yang_Koba:13}
%S. Yang, M. Kobayashi, D. Gesbert, and X. Yi, ``Degrees of freedom of time correlated MISO broadcast channel with delayed CSIT,'' {\em IEEE Transactions on Information Theory}, vol. 59, no. 1, pp. 315-328, Jan. 2013.
%
%
%\bibitem{Jafar_Index_coding}
%S. A. Jafar,
%\newblock ``Topological interference management through index coding,''
%\newblock {\em IEEE Transactions on Information Theory,}  vol. 60, no. 1, pp. 529-568, Jan. 2014.
%
%
%
%\bibitem{C-RAN}
%China Mobile Research Institute,
%\newblock ``C-RAN: The road toward Green RAN,''
%\newblock {\em White Paper,} Oct.
%2011. [Online:] http://labs.chinamobile.com/report/view59826.
%
%
%
%\bibitem{Andrews_Tractable}
%J. G. Andrews, F. Baccelli, and R. K. Ganti,
%\newblock ``A tractable approach to coverage and rate in cellular networks,''
%\newblock {\em IEEE Transactions on Communications,} vol. 59, no. 11, pp. 3122-3134, Nov. 2011.
%
%
%
%
%
%\bibitem{Gesbert}
% D. Gesbert, S. Hanly, H. Huang, S. Shamai, O. Simeone, and W. Yu,
%\newblock ``Multi-cell MIMO cooperative networks: A new look at interference,''
%\newblock {\em IEEE Journal on Sel. Areas in Communications,} vol. 28, no. 9, pp. 1380
% - 1408, Dec. 2010.
%
%
%\bibitem{3gpp}
%3rd Generation Partnership Project,
%\newblock ``Technical Specification Group Radio Access Networks; Deployment aspects (Release 7),''
%\newblock {\em3GPP TR 25.943 V7.0.0.}


\end{thebibliography}
\end{document}